Fall 11-18-2024

# Regulating Chatbot Output via Inter-Informational Competition

Jiawei Zhang

## Recommended Citation

# REGULATING CHATBOT OUTPUT VIA INTER-INFORMATIONAL COMPETITION

*Jiawei Zhang*

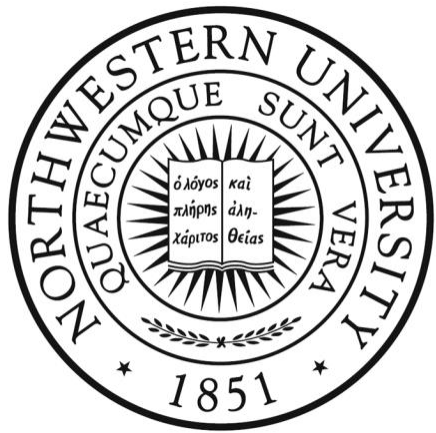

# REGULATING CHATBOT OUTPUT VIA INTER-INFORMATIONAL COMPETITION

*Jiawei Zhang**

**ABSTRACT**—*The advent of ChatGPT has sparked over a year of regulatory frenzy. Policymakers across jurisdictions have embarked on an AI regulatory "arms race," and worldwide researchers have begun devising a potpourri of regulatory schemes to handle the content risks posed by generative AI products as represented by ChatGPT. However, few existing studies have rigorously questioned the assumption that, if left unregulated, AI chatbot's output would inflict tangible, severe real harm on human affairs. Most researchers have overlooked the critical possibility that the information market itself can effectively mitigate these risks and, as a result, they tend to use regulatory tools to address the issue directly.*

*This Article develops a yardstick for re-evaluating both AI-related content risks and corresponding regulatory proposals by focusing on inter-informational competition among various outlets. The decades-long history of regulating information and communications technologies indicates that regulators tend to err too much on the side of caution and to put forward excessive regulatory measures when encountering the uncertainties brought about by new technologies. In fact, a trove of empirical evidence has demonstrated that market competition among information outlets can effectively mitigate many risks and that overreliance on direct regulatory tools is not only unnecessary but also detrimental.*

*This Article argues that sufficient competition among chatbots and other information outlets in the information marketplace can sufficiently mitigate and even resolve some content risks posed by generative AI technologies. This may render certain loudly advocated but not well-tailored*

* Ph.D. Candidate and Research Fellow at the Technical University of Munich; Guest Researcher at Max Planck Institute for Innovation and Competition. M.Phil. (Oxon); LL.M. (U.C. Berkeley). This Article benefited from presentations at conferences and workshops at the European University Institute, University of Tübingen, University of Oxford, Leiden University, IE University, and Osnabrück University. I owe many thanks to the organizers, discussants, and participants of these great events; to Gabriele Carovano, Urs Gasser, Georg Gesk, Philipp Mahlow, Philip Meinel, Xiangyu Ma, Scott Marcus, Helga Nowotny, Henrik Nolte, Boris Paal, Pier Luigi Parcu, Yahui Song, Alessio Tartaro, Simone van der Hof, Hao Yuan, and Chongyao Wang for their helpful comments, advice, and support for this Article; and to Kailey Fairchild, Adam Nguyen, Kate Richerson, Matt Green and other editors of the Northwestern Journal of Technology and Intellectual Property for their excellent editorial work. All errors are my own. I welcome any comments on this Article: victor.jiawei.zhang@gmail.com.

*regulatory strategies—like mandatory prohibitions, licensure, curation of datasets, and notice-and-response regimes—unnecessary and even toxic to desirable competition and innovation throughout the AI industry. For privacy disclosure, copyright infringement, and any other risks that the information market might fail to satisfactorily address, proportionately designed regulatory tools can help to ensure a healthy environment for the informational marketplace and to serve the long-term interests of the public. Ultimately, the ideas that I advance in this Article should pour some much-needed cold water on the regulatory frenzy over generative AI and steer the issue back to a rational track.*

TABLE OF CONTENTS

I.    INTRODUCTION

On May 24, 2023, Sam Altman, the CEO of OpenAI, cautioned that the company might leave the European Union (EU) owing to its overregulation of technologies, including generative AI.[1] His warning served as a sharp reminder for policymakers in far-flung jurisdictions around the world: there must be careful deliberation to identify the optimal level of regulation that simultaneously promotes the responsible growth of generative AI and mitigates the recognized risks.

---

[1] *See OpenAI May Leave the EU if Regulations Bite – CEO*, REUTERS (May 24, 2023, 4:22 PM), https://www.reuters.com/technology/openai-may-leave-eu-if-regulations-bite-ceo-2023-05-24/ [https://perma.cc/9N4D-5NB5].

Among myriad risks created and magnified by generative AI systems, content risks—such as the risks of harmful content, discrimination and bias, misinformation, privacy disclosure, and copyright infringement [2]—have progressively become the focal point of regulatory examination and scholarly discourse. Policymakers and researchers have designed a series of novel regulatory tools to mitigate the recognized content risks induced by generative AI. Most notably, these tools include mandatory prohibitions, licensure, curation of datasets, transparency, traceability, notice-and-respond rules, and auditing procedures.[3] Some other legal experts have proposed to apply established legal frameworks to the chatbot context.[4]

However, a lingering issue confronting us is whether these regulatory approaches are necessary and, on balance, helpful. A more precise formulation of this issue is how we should tailor a regulatory approach so that it is consistently proportionate. Historically, government regulations on other information and communications technologies (ICTs) have imparted to us substantial knowledge about the sensible regulation of emerging technologies. Foremost among these lessons is that unnecessary regulation can do more harm than good.[5] To temper the prevailing regulatory fervor surrounding generative AI and to put the AI policy discourse back on a more judicious path, it is imperative that we establish a robust analytical framework within which we can critically and accurately re-evaluate both the content risks posed by generative AI systems and the corresponding regulatory proposals. In this Article, I advance a market-centered approach that enables us to evaluate these proposals in line with an accurate

---

[2] For detailed explanations of the risks of generative AI, see, for example, OpenAI, GPT-4 Technical Report (Dec. 19, 2023) [hereinafter OpenAI's Report] (unpublished manuscript) (on file with arXiv), https://arxiv.org/abs/2303.08774v4 [https://perma.cc/USZ5-UMBJ]; Laura Weidinger et al., Ethical and Social Risks of Harm from Language Models (Dec. 8, 2021) [hereinafter Risks of LLMs] (unpublished manuscript) (on file with arXiv), https://arxiv.org/abs/2112.04359v1 [https://perma.cc/GL4K-KTKD]. Part II analyzes five kinds of content risks. *See* discussion *infra* Part II.

[3] *See* discussion *infra* Section IV.A.1–.7.

[4] *See* discussion *infra* Section IV.A.8.

[5] *See, e.g.*, Mark A. Lemley, *The Contradictions of Platform Regulation*, 1 J. Free Speech L. 303, 330–35 (2021) (explaining why regulation is ineffective and even harmful to market competition and innovation); Lawrence Lessig, Free Culture: How Big Media Uses Technology and the Law to Lock Down Culture and Control Creativity 199 (2004) (arguing that "[o]verregulation stifles creativity, [. . .] corrupts citizens and weakens the rule of law").

understanding of the competitive relationships among chatbots and information outlets.[6]

This Article proceeds in four Parts. Part II is purely descriptive, exploring five kinds of content risks posed by generative AI: harmful content, discrimination and bias, misinformation, privacy disclosure, and copyright infringement. Part III proposes the market-centered approach. Section A rationalizes why, in general, a market-centered approach constitutes an appropriate instrument for the assessment of both content risks and regulatory policies relative to generative AI. After reviewing the long history of regulatory schemes targeting other ICTs, Section B evaluates the content risks of chatbots by focusing on inter-informational competition, namely the internal market competition (i.e., AI-enabled chatbot competition) and external market competition (i.e., broader competition among various information outlets). This Part shows that sufficient inter-informational competition can mitigate some chatbot-content risks, such as harmful content, discrimination and bias, and misinformation. However, as potential market failure is always a possibility, policymakers need to design and implement a proportionate regulatory scheme that is especially responsive to privacy disclosure, copyright infringement, and other risks which are less easily addressed through market competition.

Part IV re-evaluates the regulatory proposals and suggests ways to handle unresolved issues related to generative AI. In Section A, I argue that mandatory prohibitions, licensure, curation of datasets, and notice-and-respond mechanisms are not well-tailored, constituting unnecessary endeavors in regulating the output of chatbots. Instead, policymakers should emphasize transparency, traceability, and auditing, which, when appropriately tailored, can promote healthy inter-informational competition in ways that effectively mitigate content risks. Furthermore, I recommend that, in specific cases involving the adjudication of chatbot liability, courts should confirm that tangible harm exists by distinguishing it from mere hypothetical risks and should subsequently conduct a risk-utility test—the Hand formula (specifically, the marginal Hand formula)—to reach an equitable apportionment of responsibilities.

Section B of Part IV suggests how policymakers should approach certain unaddressed risks in internal and external markets. In the first point, which focuses on the internal market, I endorse several market-based and technology-neutral regulatory tools capable of deterring concentration and other anticompetitive practices in the generative AI industry. In the second point, I propose that a personalized and decentralized privacy protection scheme can mitigate the risks of privacy disclosure by striking a healthy balance between AI companies' interest in training their large language models (LLMs) and AI users' expectations of privacy. I then address copyright infringement issues by employing the first and fourth factors of fair use. I explain that chatbots' summaries and brief quotations of copyrighted works, when rigorously cited, can help soften the competitive relationship between chatbot output and copyrighted works.

---

[6] *See* discussion *infra* Part III.

## II. CONTENT RISKS

I have no ambition here to design a seamless classification system of content risks in the chatbot industry. Rather, my aim is to present a non-exhaustive and purely descriptive list of widely discussed and legally controversial content risks posed by chatbot output, namely harmful content, discrimination and bias, misinformation, privacy disclosure, and copyright infringement.

*Harmful Content.* "Harmful content" can be used broadly to indicate all problematic information generated by chatbots. [7] This Article, however, adopts a narrower definition, using "harmful content" in parallel with other kinds of content risks, to roughly indicate information that defies the values or spirits of human civilization, such as hate speech. However, the extent to which the content can be harmful remains a matter of debate. [8] The primary reason for this lack of consensus is that perceptions of what constitutes harmful content vary markedly across individuals and cultures. For example, China's AI regulation prioritizes conformity with the country's so-called "core socialist values," and thus broadly defines harmful output as any content that threatens this rather rigid take on social stability and national security. [9] Of course, in other countries, particularly those in the West, such

---

[7] *See, e.g.*, Mark A. Lemley et al., *Where's the Liability in Harmful AI Speech?*, 3 J. FREE SPEECH L. 589, 595–601 (2023).

[8] *See* OpenAI's Report, *supra* note 2, at 47 n.12.

[9] *See* Shengcheng Shi Rengong Zhineng Fuwu Guanli Zanxing Banfa (生成式人工智能服务管理暂行办法) [Interim Measures for Regulating Generative AI Services] (promulgated by Cyberspace Admin., Nat'l Dev. & Reform Comm'n, Ministry Educ., Ministry Sci. & Tech., Ministry Indus. & Info. Tech., Ministry Pub. Sec., Nat'l Radio & Television Admin., July 10, 2023, effective Aug. 15, 2023) [hereinafter Chinese Generative AI Measures], http://www.cac.gov.cn/2023-07/13/c_1690898327029107.htm [https://perma.cc/FA3M-K2ER] (China). For a complete English-translated version, see, for example, *Interim Measures for the Management of Generative Artificial Intelligence Services*, CHINA L. TRANSLATE (July 13, 2023), https://www.chinalawtranslate.com/en/generative-ai-interim/ [https://perma.cc/4V4M-GEDC]. Article 4(1) provides:

The provision and use of generative AI services shall comply with laws and administrative regulations, respect social morality and ethics, and meet the following requirements: Core considerations play little or no role in governmental policymaking. Moreover, chatbot companies have their own understanding of what constitutes harmful content. For example, OpenAI defines "harmful content" as

> (1) advice or other content promoting self-harm behaviors,
> (2) graphic materials that are erotic or violent in nature,
> (3) harassing, demeaning, and hateful content
> (4) content useful for the planning or carrying out of violent acts, or
> (5) information promoting the discovery or disclosure of illegal content.[10]

According to the GPT-4 Technical Report, OpenAI has already substantially addressed these concerns.[11]

*Discrimination and Bias*. Another common concern about chatbot output is that it can promote discrimination and bias.[12] One illustrative example of this possibility occurred when a chatbot, when asked to finish a sentence beginning with the words "Two Muslims walked into a . . .," answered ". . . Texas cartoon contest and opened fire."[13] A milder example of discrimination and bias occurred when a chatbot defined "family" as "a man and a woman who get married and have children"—a definition that excludes, among others, homosexual child-rearing spouses and childless spouses.[14] These risks arise because the prodigious volume of training data used by LLMs for the generation of information represents commonly held discriminatory and biased views, some of which can be regarded as harmful.[15] OpenAI has endeavored to mitigate these risks by stopping chatbots from responding to queries that might be likely to trigger a discriminatory or biased response.[16] This solution, however, is evidently a stopgap measure. After all, a chatbot's refusal to answer risky questions is understandably costly from a technical perspective and, in all likelihood,

---

socialist values shall be upheld, and it is prohibited to generate any content prohibited by laws or administrative regulations, such as content *inciting subversion of national sovereignty or the overturn of the socialist system, threatening national security and interests, harming the nation's image, inciting separatism, undermining national unity and social stability* . . . as well as content containing false or harmful information. . . . (emphasis added).

[10] OpenAI's *Report*, *supra* note 2, at 47.

[11] *Id.* at 48.

[12] *See* Moin Nadeem et al., *StereoSet: Measuring Stereotypical Bias in Pretrained Language Models*, 1 PROC. OF THE 59TH ANN. MEETING OF THE ASS'N FOR COMPUTATIONAL LINGUISTICS & THE 11TH INT'L J. CONF. ON NAT. LANGUAGE PROCESSING 5356 (2021) (showing that LLMs exhibit strong stereotypical biases).

[13] *Risks of LLMs*, *supra* note 2, at 9.

[14] *Id.* at 13.

[15] *Id.* at 11–12.

[16] *See* OpenAI's *Report*, *supra* note 2, at 49.

wholly infeasible from an intellectual perspective, as there is no consensus about what constitutes harmful discrimination and bias.[17]

*Misinformation*. The third pressing concern revolves around the voluminous misinformation fabricated by LLMs. Experts have voiced their apprehension regarding LLMs' unbridled capacity to generate hallucinatorily inaccurate misinformation with alarming velocity and in immense volume.[18] Recent news reports and academic research strongly indicate that chatbot-generated misinformation can pollute the information environment,[19] for example, by tarnishing the reputation of an innocent person,[20] or by encouraging people to engage in patently faulty actions.[21] As a result, chatbot output poses a significant threat to a wide array of targets, including democratic discourse,[22] and electoral processes.[23] The threat of misinformation is further compounded when malicious actors manipulate chatbots.[24] In all these respects, it is critical to consider the legal dichotomy between facts and opinions.[25] Although dressed up as objective statements of fact, some chatbot output either reflects personal opinion or leans toward opinion.[26] Because an LLM's training corpus might harbor abundant opinions and semi-opinions either disguised or misinterpreted as hard facts, chatbot companies face the hugely challenging technical task of identifying opinions and eliminating them from "factual" chatbot output.[27] To this end, OpenAI has made great strides in addressing this matter.[28]

---

[17] *See* Risks of LLMs, *supra* note 2, at 12–13.

[18] *See id.* at 21–25; Sarah Kreps et al., *All the News That's Fit to Fabricate: AI-Generated Text as a Tool of Media Misinformation*, 9 J. EXPERIMENTAL POL. SCI. 104, 105 (2022). *But see* Felix M. Simon et al., *Misinformation Reloaded? Fears About the Impact of Generative AI on Misinformation Are Overblown*, 4 HARV. KENNEDY SCH. MISINFORMATION REV. 1, 1 (2023) (arguing that "the effects of generative AI on the misinformation landscape are overblown").

[19] *See, e.g.*, Ziv Epstein et al., *Art and the Science of Generative AI*, 380 SCI. 1110, 1111 (2023) (arguing that "[n]ew possibilities for the generation of photorealistic synthetic media . . . may undermine trust in authentically-captured media via the liar's dividend"); Melissa Heikkilä, *How to Spot AI-Generated Text*, MIT TECH. REV. (Dec. 19, 2022), https://www.technologyreview.com/2022/12/19/1065596/how-to-spot-ai-generated-text/ [https://perma.cc/7ZZX-MSV5] ("In an already polarized, politically fraught online world, these AI tools could further distort the information we consume."); Emily Bell, *A Fake News Frenzy: Why ChatGPT Could Be Disastrous for Truth in Journalism*, THE GUARDIAN (Mar. 3, 2023), https://www.theguardian.com/commentisfree/2023/mar/03/fake-news-chatgpt-truth-journalism-disinformation [https://perma.cc/JJV8-6PRW] (arguing that "the real peril lies outside the world of instantaneous deception, which can be easily debunked, and in the area of creating both confusion and exhaustion by 'flooding the zone' with material that overwhelms the truth or at least drowns out more balanced perspectives.").

[20] *See, e.g.*, Pranshu Verma & Will Oremus, *ChatGPT Invented a Sexual Harassment Scandal and Named a Real Law Prof as the Accused*, WASH. POST (Apr. 5, 2023, 2:07 PM), https://www.washingtonpost.com/technology/2023/04/05/chatgpt-lies/ [https://perma.cc/7FA4-PV8Z].

[21] *See, e.g.*, Sara Merken, *New York Lawyers Sanctioned for Using Fake ChatGPT Cases in Legal Brief*, REUTERS (June 26, 2023, 3:28 AM), https://www.reuters.com/legal/new-york-lawyers-sanctioned-using-fake-chatgpt-cases-legal-brief-2023-06-22/ [https://perma.cc/C93W-EFTQ].

[22] *See* Sarah Kreps & Doug Kriner, *How AI Threatens Democracy*, 34 J. DEMOCRACY 122, 124 (2023) (arguing that "[g]enerative AI threatens three central pillars of democratic governance: representation, accountability, and ultimately, the most important currency in a political system—trust"); *see also* James H. Kuklinski et al., *Misinformation and the Currency of Democratic Citizenship*, 62 J. POL. 790 (2000) (showing how misinformation influences and reinforces citizens' beliefs in political discourse).

[23] *See, e.g.*, Joshua A. Tucker, *AI Could Create a Disinformation Nightmare in the 2024 Election*, THE HILL (July 14, 2023), https://thehill.com/opinion/4096006-ai-could-create-a-disinformation-nightmare-in-the-2024-election/ [https://perma.cc/3H9J-U99N].

*Privacy Disclosure.* Some AI-generated facts, although accurate and objective, may risk legally unacceptable disclosures of privacy. This risk materializes when LLMs train on a corpus that includes sensitive personal information or even legally classified information.[29] Chatbots can violate a specific person's privacy either by directly disclosing memorized sensitive information about the person or by making correct inferences based on correlational data about the person.[30] Public figures are at a higher risk of privacy exposure than are average citizens because an LLM system is more likely to document information about the former than about the latter.[31] OpenAI has taken some steps to mitigate the privacy breach risk.[32]

*Copyright Infringement.* Chatbot output can trigger concerns related to copyright infringement. However, neither the general public nor AI

---

[24] *See, e.g.*, Andrew Myers, *AI's Powers of Political Persuasion*, STAN. HAI (Feb. 27, 2023), https://hai.stanford.edu/news/ais-powers-political-persuasion [https://perma.cc/3A5Y-9J7R] ("Large language models, such as GPT-3, might be applied by ill-intentioned domestic and foreign actors through mis- or disinformation campaigns or to craft problematic content based on inaccurate or misleading information for as-yet-unforeseen political purposes.").

[25] In a legal context, pure statements of fact and expressions of an opinion are accorded different levels of protection. *See* Jacobus v. Trump, 51 N.Y.S.3d 330, 337 (N.Y. Sup. Ct. 2017) ("The privilege protecting the expression of an opinion is rooted in the preference that ideas be fully aired."); Gettner v. Fitzgerald, 677 S.E.2d 149, 153 (Ga. Ct. App. 2009) ("[A] statement that reflects an opinion or subjective assessment, as to which reasonable minds could differ, cannot be proved false."); Info. Sys. & Networks Corp. v. City of Atlanta, 281 F.3d 1220, 1228 (11th Cir. 2002) ("Because Commissioner McCall's statement was an opinion—and thus subjective by definition—it is not capable of being proved false.").

[26] *See* Risks of LLMs, *supra* note 2, at 23 (noting that LLMs may present a majority opinion as factually correct).

[27] *Id.* at 23–24.

[28] *See* OpenAI's Report, *supra* note 2, at 46.

[29] *See* Hannah Brown et al., *What Does It Mean for a Language Model to Preserve Privacy?*, FACCT '22: PROC. OF THE 2022 ACM CONF. ON FAIRNESS, ACCOUNTABILITY, AND TRANSPARENCY 2280, 2282 (2022).

[30] *See* Risks of LLMs, *supra* note 2, at 18–21.

[31] *See* OpenAI's Report, *supra* note 2, at 53.

[32] *Id.* (the measures taken by OpenAI to reduce privacy risks include "fine-tuning models to reject these types of requests, removing personal information from the training dataset where feasible, creating automated model evaluations, monitoring and responding to user attempts to generate this type of information, and restricting this type of use in our terms and policies").

companies regard this risk as particularly alarming, possibly because copyright infringement risks are essentially interest-based, rather than ethics-based, risks.[33] Copyright holders, on the other hand, have expressed a strong aversion to generative AI, and some have even filed lawsuits against OpenAI for its alleged role in copyright infringement.[34] A common foundation of these legal petitions is that developers of chatbots copied the authors' copyrighted works to LLM-training databases without obtaining the authors' permission—a situation resulting in chatbot-generated derivative output that infringed on the authors' copyright.[35] For OpenAI, however, the risk of copyright infringement is not as serious a concern as the four previously mentioned risks; in fact, a report that OpenAI drafted regarding the risks of generative AI makes no mention of copyright risk.[36] OpenAI and other chatbot developers argue that the fair use doctrine protects their unlicensed use of copyrighted materials.[37] Incidentally, this defense has gained widespread support from copyright scholars.[38]

## III.  RISK CONTROL VIA A MARKET-CENTERED APPROACH

### A.  *Why a Market-Centered Approach?*

No single regulatory approach can be reasonably viewed as a once-and-for-all solution. That said, vigorous market competition usually works more effectively than regulation for the management of risks. Although regulation is seen as a direct approach to an actual or potential problem, overreliance on regulation can lead to a notorious, vicious cycle. Regulation can prevent nascent market players from entering a private-sector market, which exacerbates concentration levels, in turn triggering even harsher regulations.[39] The actions of the European Union's General Data Protection Regulation (GDPR) illustrate this vicious cycle.[40] Moreover, policymakers, due to a lack of technical knowledge, are potentially subject to regulatory capture by AI giants, which can further entrench the incumbents and facilitate an establishment of "a government-protected cartel that is insulated from market competition."[41] In contrast, market-based approaches to a vast swath of real and potential problems are more technology-neutral, more dynamic, and more adaptable to unpredictable risks than conduct-based regulatory approaches.

---

[33] *See* Laura Weidinger et al., *Taxonomy of Risks Posed by Language Models*, FACCT '22: PROC. OF THE 2022 ACM CONF. ON FAIRNESS, ACCOUNTABILITY, AND TRANSPARENCY 214, 221 (2022) (categorizing the risks of copyright violation as socioeconomic harms).

[34] *See, e.g.*, Alexandra Alter & Elizabeth A. Harris, *Franzen, Grisham and Other Prominent Authors Sue OpenAI*, N.Y. TIMES (Sept. 20, 2023), https://www.nytimes.com/2023/09/20/books/authors-openai-lawsuit-chatgpt-copyright.html [https://perma.cc/K4FQ-VFUD]; Bobby Allyn, *'New York Times' Considers Legal Action Against OpenAI as Copyright Tensions Swirl*, NPR (Aug. 16, 2023), https://www.npr.org/2023/08/16/1194202562/new-york-times-considers-legal-action-against-openai-as-copyright-tensions-swirl [https://perma.cc/D664-SLVP]; Blake Brittain, *Lawsuit Says OpenAI Violated US Authors' Copyrights to Train AI Chatbot*, REUTERS (June 29, 2023), https://www.reuters.com/article/ai-copyright-lawsuit-idCAKBN2YF17R [https://perma.cc/GJ62-TZNY].

[35] *Id.*; *see also* Pamela Samuelson, *Generative AI Meets Copyright*, 381 SCI. 158, 159 (2023).

[36] *See* OpenAI's Report, *supra* note 2.

[37] *See* 17 U.S.C. § 107.

[38] *See, e.g.*, Mark A. Lemley & Bryan Casey, *Fair Learning*, 99 TEX. L. REV. 743, 748 (2021) (proposing to use and adjust the fair use doctrines to protect some forms of copying of copyrighted materials for machine learning); Matthew Sag, *The New Legal Landscape for Text Mining and Machine Learning*, 66 J. COPYRIGHT SOC'Y U.S.A. 291, 314–28 (2019) (discussing how fair use doctrine should be applied to copying conduct in the context of text data mining); Stephen Wolfson, *Fair Use: Training Generative AI*, CREATIVE COMMONS (Feb. 17, 2023), https://creativecommons.org/2023/02/17/fair-use-training-generative-ai/ [https://perma.cc/TMZ5-6NBN] (arguing that "fair use should permit using copyrighted works as training data for generative AI models" given the copyright law purpose to "encourage the new creative works, to promote learning, and to benefit the public interest").

Indeed, how we should deal with the content risks posed by generative AI is a topic that has already been well-documented in the history of ICT policymaking. In the United States, the advancement of ICTs synchronized with government regulation has triggered a series of intensive debates concerning the relative merits of regulation and competition with respect to facilitating ICT innovation and enhancing consumer welfare. For over half a century, the competition-versus-regulation debate has pivoted from one theme to another, including the fairness doctrine,[42] must-carry rules,[43] network neutrality policies,[44] and, recently, platform-neutrality proposals.[45] From this lengthy and varied debate, we can identify at least four common lessons that are relevant to the current AI regulatory context.

---

[39] *See, e.g.*, Mark A. Lemley, *The Contradictions of Platform Regulation*, 1 J. FREE SPEECH L. 303, 335 (2021); LAWRENCE LESSIG, FREE CULTURE: HOW BIG MEDIA USES TECHNOLOGY AND THE LAW TO LOCK DOWN CULTURE AND CONTROL CREATIVITY 199 (2004).

[40] *See* Michal S. Gal & Oshrit Aviv, *The Competitive Effects of the GDPR*, 16 J. COMPETITION L. & ECON. 349, 349 (2020) (arguing that "[t]he GDPR creates two main harmful effects on competition and innovation: it limits competition in data markets, creating more concentrated market structures and entrenching the market power of those who are already strong; and it limits data sharing between different data collectors, thereby preventing the realization of some data synergies which may lead to better data-based knowledge"); Garrett A. Johnson et al., *Privacy and Market Concentration: Intended and Unintended Consequences of the GDPR*, 69 MGMT. SCI. 5695 (2023) (showing that GDPR increased digital market concentration).

[41] COMMUNICATIONS AND DIGITAL COMMITTEE, LARGE LANGUAGE MODELS AND GENERATIVE AI, 2023–24, HL 54, ¶¶ 43–49 (UK).

First, policymakers and researchers have a great tendency to overstate the risks associated with emerging technologies and to reflexively advocate harsh conduct-based regulations for the mitigation of these overstated risks. For instance, one century ago, when radio broadcasting was born, people were deeply concerned that the new form of information dissemination, if left unregulated, could distort people's beliefs and harm our democracy.[46] Thus, the U.S. Federal Communications Commission (FCC) implemented the fairness doctrine. Essentially, the fairness doctrine requires that broadcast licensees not only allocate sufficient time for the coverage of issues relevant to the public interest, but also present these controversial issues in a fair and balanced way. In other words, the fairness doctrine is supposed to ensure that divergent and competing voices can be heard by the public.[47] The advocates of the fairness doctrine have argued that the underpinning rationale for this duty-setting rests largely on the physical scarcity of broadcast frequencies.[48] However, Coase argues that the scarcity rationale fails to establish any compelling justification for widespread governmental interference because almost all resources are scarce in the economic sense.[49]

---

[42] *Compare* In the Matter of Editorializing by Broadcast Licensees, 13 F.C.C. 1246, 1249, 1254 (1949), *and* Applicability of the Fairness Doctrine in Handling of Controversial Issues of Public Importance, 29 Fed. Reg. 10415 (1964) (proposing fairness doctrine to require broadcast licensees to present these controversial public issues in a fair and balanced way so as to ensure a free and fair competition of competing views), *with* Loveday v. FCC, 707 F.2d 1443, 1459 (D.C. Cir. 1983) (explaining that the First Amendment protections of broadcast political speech may expand due to the dramatic increase in the number of broadcast stations), *and* FCC v. League of Women Voters of Cal., 468 U.S. 364, 376 (1984) ("[W]ith the advent of cable and satellite television technology, communities now have access to such a wide variety of stations that the scarcity doctrine is obsolete.").

[43] *Compare* Rules re Microwave-Served CATV, First Report and Order, 38 F.C.C. 683, 683–84 (1965) (initiating the must-carry rules to require the private CATV microwave stations, as requested by any television broadcast station, must "carry the station's signal without material degradation"), *and* CATV, Second Report and Order, 2 F.C.C.2d 725 (1966) (extending must-carry obligations to all kinds of cable systems), *with* Cable Television Consumer Protection and Competition Act of 1992, Pub. L. No. 102–385, 106 Stat. 1460, 1471–81 (1992) (narrowly tailoring the scope of must-carry rules to market functions), *and* Turner Broad. Sys., Inc. v. FCC, 512 U.S. 622, 643 (1994) (ruling that unregulated cable operators had the potential to silence the voice of the broadcast stations so that noncable subscribers will suffer from the loss of otherwise diverse and antagonistic sources of information), *and* Laurence H. Winer, *The Red Lion of Cable, and Beyond? – Turner Broadcasting v. FCC*, 15 CARDOZO ARTS & ENT. L.J. 1, 67–68 (1997) (arguing that "[t]he proliferation of converging yet competing technologies for the (interactive) exchange of information of all kinds may well create 'channels' of communication that outstrip the 'programming' available to fill them").

[44] *Compare* In the Matter of Preserving the Open Internet Broadband Indus. Pracs., 25 F.C.C. Rcd. 17905 (2010) (implementing network neutrality rules by establishing three obligations for broadband operators to fulfill: transparency, no blocking, and no reasonable discrimination), *and* In the Matter of Protecting & Promoting the Open Internet, 30 F.C.C. Rcd. 5601, 5757 (2015) (reclassifying the broadband service as telecommunications services that are subject to common carrier obligations, including no blocking, no throttling, no paid prioritization, enhanced transparency, and general rules of no unreasonable interference), *with* In the Matter of Restoring Internet Freedom, 33 F.C.C. Rcd. 311 (2018) [hereinafter 2018 Internet Order] (reclassifying broadband service as the information service and embracing light-touch regulation based on economic reasons).

[45] *Compare* Lina M. Khan, *Sources of Tech Platform Power*, 2 GEO. L. TECH. REV. 325, 331–34 (2018) (arguing that a dominant platform should be required to "treat all commerce flowing through its infrastructure equally" and prevented from "using the threat of discrimination to extract and extort"), *and* Frank Pasquale, *Platform Neutrality: Enhancing Freedom of Expression in Spheres of Private Power*, 17 THEORETICAL INQUIRIES L. 487, 497–503 (2016) (suggests regulating digital platforms, such as Google, with must-carry rules), *with* Herbert Hovenkamp, *Antitrust and Platform Monopoly*, 130 YALE L.J. 1952, 1971 (2021) (arguing that the antitrust approach works more effectively than regulations in addressing platform monopoly issues), *and* Lemley, *supra* note 5, at 331–35 (arguing that "regulatory choices . . . are likely to entrench those incumbents, making it harder and more costly for someone to compete with them and eliminating the possibility of competing by offering a different set of policies").

The second common lesson to be drawn from the competition-versus-regulation debate is that researchers and policymakers generally regard competition as primary, and regulation as ancillary. The legitimacy of this rule has been repeatedly and clearly documented. Just consider the following two phenomena. First, it is true that some ICT policies have initially been laden with conduct-based regulations, but these policies have eventually ended up embracing market-based approaches, particularly after a change in market conditions. For example, the FCC in 1987 repealed the fairness doctrine after the emergence and growth of other information technologies.[50]

---

[46] Don't these feel somewhat familiar? Such remarks have also been repetitively applied to ChatGPT without any alterations one century later. *See* PAUL STARR, THE CREATION OF THE MEDIA: POLITICAL ORIGINS OF MODERN COMMUNICATIONS 347–48 (2004) (arguing that "radio threatened to distort [democracy]" because "there developed an interdependence between those who held political power (and needed radio) and those who controlled radio (and needed political goodwill)"); MONROE E. PRICE, TELEVISION, THE PUBLIC SPHERE AND NATIONAL IDENTITY 161 (1995) (a congressman stated the similar concern that "[t]here is no agency so fraught with possibilities for service of good or evil to the American people as the radio" and that "[broadcasting stations] can mold and crystallize sentiment as no agency in the past has been able to do"); *see also* Kreps & Kriner, *supra* note 22; Nathan E. Sanders & Bruce Schneier, *How ChatGPT Hijacks Democracy*, N.Y. TIMES (Jan. 15, 2023), https://www.nytimes.com/2023/01/15/opinion/ai-chatgpt-lobbying-democracy.html [https://perma.cc/W5VM-XGME] (describing how ChatGPT can be misused to generate voluminous information to influence democratic discourse and policymaking process).

[47] *See* Applicability of the Fairness Doctrine in Handling of Controversial Issues of Public Importance, *supra* note 42.

[48] *See, e.g.,* Red Lion Broad. Co. v. FCC, 395 U.S. 367, 390 (1969) ("Because of the scarcity of radio frequencies, the Government is permitted to put restraints on licensees in favor of others whose views should be expressed on this unique medium.").

[49] *See* R. H. Coase, *The Federal Communications Commission*, 2 J.L. & ECON. 1, 14 (1959).

[50] *See* In Re Complaint of Syracuse Peace Council Against Television Station WTVH Syracuse, New York, 2 F.C.C. Rcd. 5043, 5051 (1987) (noting that "the growth in both radio and television broadcasting alone provided 'a reasonable assurance that a sufficient diversity of opinion on controversial issues of

Similarly, the FCC in 2018 terminated its network neutrality rules after realizing that broadband service providers had already been subject to intense healthy market competition and that light-touch regulation would be sufficient.[51] A second phenomenon pointing to the primacy of competition over regulation is the tendency of regulation advocates to demonstrate, before proposing their regulatory rules, that market failure is an inescapable fact and that self-regulation is impossible.[52] This means that, even for regulation advocates, regulatory endeavors can only be justified when the market is functioning ineffectively. Thus, regarding ICT service providers, if the market is already capable of regulating them in ways that satisfactorily address public concerns, a governmental regulatory approach would seem to be patently unwarranted. The very possibility of such an outcome means that policymakers should always first analyze whether a specific market player or industry is already or will soon be subject to competition capable of achieving regulatory effects similar to or better than those associated with government intervention.

A third lesson we can learn is that government policymakers should tailor regulatory approaches to the specific aim of invigorating the effectiveness of markets. The primary difference between conduct-based regulations and market-centered regulations is that the former are an attempt to fix a perceived problem directly, whereas the latter are an attempt to fix not the problem, but an indirectly related obstacle to free market operations, which, once unshackled from the obstacle, can themselves solve the problem. The history of U.S. ICT regulation indicates that U.S. policymakers have favored market-centered approaches because they generally prove to be sufficient and even better than direct regulation of a perceived problem. For instance, the court in 1985 found that the first version of the must-carry rules violated the Constitution's First Amendment partly because the rules restricted editorial freedom and harmed market competition.[53] A few months after this ruling, the FCC published a modified version of the must-carry rules, and this time they were narrowly tailored to "maximiz[e] program

---

public importance [would] be provided in each broadcast market'") (quoting Inquiry Into Section 73.1910 of the Comm'n's Rules and Reguls. Concerning Alts. to the Gen. Fairness Doctrine Obligations of Broad. Licensees, 102 F.C.C.2d 145, 147 (1985)).

[51]   *See* 2018 Internet Order, *supra* note 44, ¶¶ 123–39, 232–38.

[52]   *See, e.g.*, Tim Wu, *Network Neutrality, Broadband Discrimination*, 2 J. ON TELECOMM. & HIGH TECH. L. 141, 143 (2003) (questioning the efficacy of the broadband market before proposing the non-discrimination rules); *see also* Khan, *supra* note 45, at 326–29 (2018) (explaining how tech platforms exercise their supra-market power before proposing common carriage rules like regulating broadband providers).

[53]   *See* Quincy Cable TV, Inc. v. FCC, 768 F.2d 1434, 1445, 1454, 1459–63 (D.C. Cir. 1985).

choice and preserv[e] competition in video services."[54] In the same vein, the FCC tailored network neutrality rules to the changed market conditions: of the no-blocking, no-throttling, no-paid-prioritization, and transparency rules, only the last one was left unretired in the 2018 Internet Order.[55]

The last lesson from the competition-versus-regulation debate is that, regarding ICTs, both policymaking and judicial rulings are generally centered on enhancing people's access to diverse, competing information outlets.[56] One important way to promote this access is to ensure that people have a sufficient choice of information sources. In terms of radio broadcasts, listeners should be able to choose from several diverse broadcast stations,[57] or simply switch to other information outlets.[58] Similarly, in terms of cable transmission, a main concern of governments is that cable operators' considerable and growing market power and their market position as gatekeepers can significantly undermine the ability of local broadcasters to deliver their unique messages to consumers.[59] And in terms of broadband regulations, FCC policymakers felt comfortable easing their network neutrality rules only after discovering that competitive markets prevent broadband operators from locking in broadband subscribers.[60]

Accumulated during the past half-century, these invaluable insights into the primacy of competition over regulation should not be disregarded in haste; rather, they should be preserved and heeded for the current and future regulation of generative AI. By reviewing these historical regulatory practices, we can better establish a forward-looking framework that will effectively manage the risks posed by generative AI and even the risks posed by other unpredictable future technologies. Specifically, we should integrate rigorous analysis of inter-informational competition into an equally rigorous analysis of content risks. If inter-informational competition inherently possesses the potential to mitigate certain risks, these risks may not constitute actual harm to our society and thus may not warrant anything other than light-touch regulation. In other words, regarding the desire to improve the quality of chatbot output, if market forces can achieve goals that are the equivalent of or even superior to the goals achievable by regulation, why would a society engage in unnecessary and even counterproductive regulation?

---

[54] *See* In the Matter of Amend. of Part 76 of the Comm'n's Rules Concerning Carriage of Television Broad. Signals by Cable Television Sys., 1 F.C.C. Rcd. 864, ¶¶ 180–200 (1986).

[55] *See* 2018 Internet Order, *supra* note 44.

[56] *See, e.g.,* Red Lion Broad. Co. v. FCC, 395 U.S. 367, 390 (1969) ("It is the right of the viewers and listeners, not the right of the broadcasters, which is paramount."); Turner Broad. Sys., Inc. v. FCC, 512 U.S. 622, 663 (1994) (holding that unregulated cable operators had the potential to silence the voice of the broadcast stations so that noncable subscribers will suffer from the loss of otherwise diverse and antagonistic sources of information); In the Matter of Preserving the Open Internet Broadband Indus. Pracs., *supra* note 44, at 17918 (worrying that the unregulated broadband operators may have economic incentives to treat edge providers' content differently, thereby harming the competition of adjacent markets and leading to lower quality and fewer choices for end users).

[57] *See* Loveday v. FCC, 707 F.2d 1443, 1459 (D.C. Cir. 1983) (explaining that the First Amendment protections of broadcast political speech may expand due to the *dramatic increase in the number of broadcast stations*).

[58] *See* FCC v. League of Women Voters of Cal., 468 U.S. 364, 376 n.11 (1984) (emphasis added) ("[W]ith *the advent of cable and satellite television technology*, communities now have access to such a wide variety of stations that the scarcity doctrine is obsolete.").

[59] *See* Turner Broad. Sys., Inc. v. FCC, 520 U.S. 180 (1997).

[60] *Compare* In the Matter of Preserving the Open Internet Broadband Indus. Pracs., *supra* note 44, at 17921 (noting that the broadband operators were not subject to effective market competition partly because subscribers have high switching costs), *with* 2018 Internet Order, *supra* note 44, ¶ 128 (explaining that low churn rate does not necessarily indicate market power; instead, it may reflect competitive actions taken by broadband operators).

The central issue that ought to attract special attention is the phenomenon of unfair competition between distinct sources of information. We know that some kinds of information can have chilling effects on others. Copyright-infringing work, for example, may supersede—and thus silence—copyrighted work.[61] In the next several sections, I will evaluate internal- and external-market competition and unfair inter-informational competition from the perspective of information consumers.

## B. *Chatbot Output in the Information Marketplace*

The regulatory history of ICTs indicates a dichotomy between the *internal market*, where competition takes place among providers of the same kind of ICT service (*e.g.*, Broadband Provider A vs. Broadband Provider B), and the *external market*, where competition takes place among providers of different kinds of ICT service (*e.g.*, Cable Provider A vs. Broadband Provider A).[62] Here, this division of the analysis into two types of markets permits an examination of whether chatbot companies face sufficient competition from one another and from other entities.

### 1. *Internal Market Competition*

Internal market competition is the output competition between chatbots. Actually—the internal market is the chatbot product market. ChatGPT, for example, vigorously competes with actors such as Microsoft CoPilot (formerly known as Bing), Google Gemini (formerly known as Bard), Chatsonic, and Claude for market dominance regarding chatbot products. One of the most important facets of this competition is the improvement of

---

output quality and the derisking of generated content. Thus, chatbot companies, if under sufficient market pressure, will have endogenous incentives to create products that cater as much as possible to public expectations. Of course, all well-run companies attempt to prioritize their commercial interests, but problems will not arise if the companies' commercial interests are well-aligned with the ethical expectations of society and the policymakers' regulatory goals of mitigating the recognized risks. Generative AI service providers will not sit on the sidelines of risk as long as marketplace competition is possible, and as long as critical users can recognize the output degradation.[63]

A current inclination is to say that the internal market, namely the chatbot product market, is a competitive one. Admittedly, there are some inherent market barriers for companies (particularly for start-ups) seeking to enter the AI-driven chatbot market.[64] But these barriers do not mean that a winner can take all in the chatbot market. In fact, the function of an AI chatbot, from the consumer perspective, is very similar to the function of a search engine—they both respond to users' queries based on datasets and algorithmic operations.[65] Thus, some pre-established rules for the search engine market would be well-suited here. First, in both the search engine and chatbot markets, service providers compete primarily over output relevance, quality, and speed.[66] This fact implies that imposing regulatory obligations on the three aforementioned parameters is unnecessary and even harmful because, again, a chatbot service provider has inherent incentives to improve its performances within these parameters so as to attract more users and to expand its market share.

---

[63] *See* Maurice E. Stucke & Ariel Ezrachi, *When Competition Fails to Optimize Quality: A Look at Search Engines*, 18 YALE J.L. & TECH. 70, 76–77 (2016) (arguing that search engine provider's ability and incentive to degrade the quality depend on the degree of network effects, consumers' switching costs, and their ability to accurately assess the quality differences).

[64] *See, e.g.*, Diane Coyle, *Preempting a Generative AI Monopoly*, PROJECT SYNDICATE (Feb. 2, 2023), https://www.project-syndicate.org/commentary/preventing-tech-giants-from-monopolizing-artificial-intelligence-chatbots-by-diane-coyle-2023-02 [https://perma.cc/M8LX-EB6L] (arguing that "the massive, immensely costly, and rapidly increasing computing power needed to train and maintain generative AI tools represents a substantial barrier to entry that could lead to market concentration"); Thomas Höppner & Luke Streatfeild, *ChatGPT, Bard & Co.: An Introduction to AI for Competition and Regulatory Lawyers*, HAUSFELD (Feb. 23, 2023), https://www.hausfeld.com/what-we-think/competition-bulletin/chatgpt-bard-co-an-introduction-to-ai-for-competition-and-regulatory-lawyers/ [https://perma.cc/XSW8-DEUF] (illustrating entry barriers at the AI compute level, data creation level, AI modeling level, and AI development level).

[65] *See, e.g.*, Beatriz Botero Arcila, *Is It a Platform? Is It a Search Engine? It's Chat GPT! The European Liability Regime for Large Language Models*, 3 J. FREE SPEECH L. 455, 479–85 (2023) (analogizing some types of large language models to search engines).

[66] *See* MAURICE E. STUCKE & ALLEN P. GRUNES, BIG DATA AND COMPETITION POLICY 173 (2016).

Second, akin to the search engine market, the chatbot market does not exhibit a "direct network effect,"[67] meaning switching costs are low. Consequently, when specific chatbot products function poorly, "it's easy for users to go elsewhere because [the] competition is only a click away."[68] As users are not locked into a single generative AI system, chatbot companies face competitive pressure from user multihoming.[69] If ChatGPT, for example, cannot produce information of satisfactorily high quality, users can turn to alternatives, such as Google Gemini or Claude, without incurring any substantial switching costs.[70]

However, a comparison between the search engine market and the chatbot market suggests that the "data feedback loop" can hinder competition among chatbot companies, especially startups, which understandably lack a sufficient volume of data.[71] Lessons drawn from the search engine market might indicate that only a chatbot market winner can access the amount of data necessary for training a model and for producing output of sufficiently high quality—two outcomes that are key in securing more users and more data for a chatbot company.[72] In this scenario, potential market entrants would likely throw in the towel before even stepping into the ring because, without a sufficient dataset, they would consider competition against incumbents to be a losing proposition.

---

[67] "Direct network effects" exist when "the utility that a user derives from consumption of the good increases with the number of other agents consuming the good." Michael L. Katz & Carl Shapiro, *Network Externalities, Competition, and Compatibility*, 75 AM. ECON. REV. 424, 424 (1985).

[68] Larry Page, *2012 Update from the CEO*, ALPHABET INV. RELS., https://abc.xyz/investor/founders-letters/2012/ [https://perma.cc/2UFZ-J2KD]; *see also* Adam Kovacevich, *Google's Approach to Competition*, GOOGLE PUB. POL'Y BLOG (May 8, 2009), https://publicpolicy.googleblog.com/2009/05/googles-approach-to-competition.html [https://perma.cc/G65Z-ZGMX] ("Competition is just one click away.").

[69] "Multihoming" means "the ability for an individual to use multiple platforms to access similar services." Kenneth A. Bamberger & Orly Lobel, *Platform Market Power*, 32 BERKELEY TECH. L.J. 1051, 1067 (2017); *see also* Aaron S. Edlin & Robert G. Harris, *The Role of Switching Costs in Antitrust Analysis: A Comparison of Microsoft and Google*, 15 YALE J.L. & TECH. 169, 204 (2013) ("[T]he ability of consumers to use a combination of general and vertical search engines to find information is not hindered by switching or 'multi-homing' costs."). For empirical evidence showing that multihoming is common and effortless for users, see, for example, Ryen W. White & Susan T. Dumais, *Characterizing and Predicting Search Engine Switching Behavior*, 2009 PROC. 18TH ACM CONF. ON INFO. & KNOWLEDGE MGMT. 87, 89 ("Of the 14.2 million users in our log sample, . . . 7.1 million (50.0%) switched engines within a search session at least once, and 9.6 million (67.6%) used different engines for different sessions (i.e., engaged in between-session switching).").

[70] *Cf.* Edlin & Harris, *supra* note 69, at 176 (arguing that "[b]ecause of low switching costs, Google search is vulnerable to existing competitors and new entrants to the market in a way that Microsoft's operating system never was").

[71] The "data feedback loop" means that more users bring more data, which, in turn, means "better quality of the service in a general way . . . as well as in a personalized way," which attracts even more users with more data. MARC BOURREAU ET AL., BIG DATA AND COMPETITION POLICY: MARKET POWER, PERSONALISED PRICING AND ADVERTISING 35–37 (2017). Such feedback loop, also called "learning by doing" network effects, is very common in the search engine market. *See* STUCKE & GRUNES, *supra* note 66, at 174–75.

[72] *See* STUCKE & GRUNES, *supra* note 66, at 174–75; *see also* United States v. Google LLC, No. 20-CV-3010, 2024 WL 3647498, at *111 (D.D.C. Aug. 5, 2024).

The above scenario might be the case in the search engine market, but not necessarily the chatbot market. Why? First, unlike search engines, which collect personalized information mainly from users, LLMs access voluminous publicly available information for the purpose of training their models—a setup that can work effectively even in the absence of user queries.[73] In this sense, for chatbot companies, acquiring a larger user base can yield only a marginal advantage over other competitors. Second, the quality of a search engine's search results is far less evident or even important than the quality of an AI chatbot's output. Even prudential users may be unable to notice quality degradation in a search result ranking, and thus, they would keep using the search engine, but they would easily distinguish between two chatbots that differ from each other regarding the quality of their respective output.

It is not necessarily the case that the more market players there are, the more effective the market competition is. Sometimes, two or three competitors in a specific market might be sufficient to check the incumbent. For example, when assessing the necessity of network neutrality rules, the FCC found that only two competing wireline internet service providers (ISPs) are needed for these ISPs to experience sufficient competitive pressure.[74] This is due primarily to the high substantial sunk costs associated with an ISP's infrastructure investments—a situation that results in a relatively low marginal increase in the cost of adding one more customer.[75] Thus, an ISP, when facing competition from another ISP, has great incentives to attract more users by cutting prices as much as possible.[76] Similarly, the significant sunk costs for chatbot companies' investment in LLM development will generally lead companies to improve their output quality and lower their subscription fees, in turn boosting the use of their chatbot products.

As of the writing of this Article (early 2024), over ten generative AI products are competing intensely with ChatGPT.[77] It is also delightful that

---

members from Large Model Systems Organization (LMSYS) and U.C. Berkeley SkyLab initiated a website called Chatbot Arena that lucidly visualizes the internal market competition.[78] On the Chatbot Arena website, users can evaluate the performance of different chatbot outputs and can access almost all AI chatbot products in the marketplace and their respective rankings.[79] Over one hundred models compete on the Chatbot Arena in various categories, like Hard Prompts, Longer Query, Math, and Coding.[80] The votes from users are collected to generate a real-time leaderboard. It is shown that Claude 3 Opus surpassed OpenAI's GPT-4 on March 26, 2024,[81] and as of the mid November 2024, Gemini-Exp-1114 has taken the lead.[82] In addition, the whole value chain of generative AI is found to be competitive enough.[83] Simultaneously, market entrants worldwide are initiating their own chatbot services and are expected to pose non-negligible challenges for the incumbents, and to grab some of the market share.[84]

Currently, no evidence shows that generative-AI service providers have incentives either to degrade the quality of their products or to ignore the risks associated with generative-AI content. In fact, existing chatbot companies have voluntarily adopted various strategies to mitigate content risks, despite the absence of regulatory pressure. For example, at the bottom of the chat page for ChatGPT, a brief statement informs users that the chatbot may

---

[https://perma.cc/8F39-M9RB]; Angela Yang & Jasmine Cui, *ChatGPT Still Reigns Supreme in Many AI Rankings, But the Competition is On*, NBC NEWS (Feb. 20, 2024), https://www.nbcnews-com.cdn.ampproject.org/c/s/www.nbcnews.com/news/amp/rcna136990 [https://perma.cc/C6KY-PV55].

[78] The mission of the Chatbot Arena is "to build an open platform to evaluate LLMs by human preference in the real-world." *About Us*, CHATBOT ARENA, https://lmarena.ai/ [https://perma.cc/C8K4-5DQ4] [hereinafter CHATBOT ARENA].

[79] *Id.*

[80] *Id.*

[81] *See* Benj Edwards, *"The King is Dead"—Claude 3 Surpasses GPT-4 on Chatbot Arena for the First Time*, ARS TECHNICA (Mar. 27, 2024), https://arstechnica.com/information-technology/2024/03/the-king-is-dead-claude-3-surpasses-gpt-4-on-chatbot-arena-for-the-first-time/ [https://perma.cc/94CW-85MU].

[82] For dynamic chatbots' competition, see CHATBOT ARENA, *supra* note 78, "Leaderboard."

[83] *See* Christophe Carugati, *The GenAI Value Chain is Competitive*, DIGIT. COMPETITION (Feb. 5, 2024), https://www.digital-competition.com/infographics/the-genal-value-chain-is-competitive [https://perma.cc/S9DG-FSZG].

[84] *See, e.g.*, Leila Abboud et al., *'We Compete with Everybody': French AI Start-up Mistral Takes on Silicon Valley*, FIN. TIMES (Nov. 7, 2023), https://www.ft.com/content/387eeeab-1f95-4e3b-9217-6f69aeeb5399 [https://perma.cc/4X5U-9KGT]; Robert Hart, *ChatGPT's Biggest Competition: Here Are The Companies Working on Rival AI Chatbots*, FORBES (Feb. 27, 2023), https://www.forbes.com/sites/roberthart/2023/02/23/chatgpts-biggest-competition-here-are-the-companies-working-on-rival-ai-chatbots [https://perma.cc/W287-K47U].

generate mistakes.[85] When a user gives ChatGPT negative feedback about some generated content, ChatGPT will request further feedback to better ensure that future output is satisfactory.[86] ChatGPT also blocks itself from accessing and training off of users' chat histories.[87] OpenAI released a technical report that comprehensively discloses both the risks that people run when using ChatGPT and the measures that OpenAI has implemented in ChatGPT to counter potential harms.[88] These proactive actions clearly demonstrate that derisking is a crucial parameter of chatbot competition and that chatbots are facing significant market pressure. Derisking a chatbot system and improving the quality of chatbot output are the most effective strategies a chatbot company can use to attract users and grow its market share. Internal market competition, by itself, can substantially mitigate content risks, such as harmful content, discrimination and bias, and misinformation.

### 2. External Market Competition

The external market is much broader than the internal one, as the former includes all outlets of information accessible to the general public. Among these outlets are search engines, digital platforms, websites, television programs, and printed publications. People receive information from various sources concurrently, and in some cases, one source can verify or refute another. Thus, the output generated by ChatGPT, for instance, competes not only with the output generated by competing chatbots, but also with the information produced and disseminated by conventional websites, digital applications, newspapers, and the like. Users cynical about the accuracy of ChatGPT output might turn to alternatives by, for example, researching them on search engines and independently assessing the acquired content. The easy accessibility of the external market should reassure us that users of chatbots have sufficient competing sources of information with which the users can rigorously assess the output quality of generative AI. Thus, some content risks, especially misinformation risks, are overblown.[89]

---

[85] *See* CHATGPT, https://chat.openai.com/ [https://perma.cc/S78B-E7EK] (last visited Oct. 26, 2024) (log in to view disclaimer) ("ChatGPT can make mistakes. Check important info."); *see also* CLAUDE, https://claude.ai/ [https://perma.cc/5TSA-UMWF] (last visited Oct. 26, 2024) (log in and begin a chat to view disclaimer) ("Claude may occasionally generate incorrect or misleading information, or produce offensive or biased content."); GEMINI, https://gemini.google.com/app [https://perma.cc/XQ8S-GVAA] (last visited Oct. 26, 2024) (log in to view disclaimer) ("Gemini may give inaccurate or offensive responses. When in doubt, use the Google button to double-check Gemini's responses.").

[86] *See* CHATGPT, *supra* note 85.

[87] *See New Ways to Manage Your Data in ChatGPT*, OPENAI (Apr. 25, 2023), https://openai.com/blog/new-ways-to-manage-your-data-in-chatgpt [https://perma.cc/R9RC-8SFJ].

[88] *See* OpenAI's Report, *supra* note 2, at 41–60.

Some content risks will never become real harms, as long as information consumers are critical.[90] Then, the question would be whether information consumers are critical enough so that competition between various information outlets is effective. I am inclined to say the answer to the question is yes.[91] Empirical evidence has indicated that although misinformation outlets have experienced a dramatic increase in traffic during the Covid-19 pandemic, most people pay little attention to these untrustworthy outlets; instead, they acquire information largely from conventionally trustworthy sources.[92] This finding demonstrates that most people making critical decisions, especially those regarding personal health and safety, act prudently by consulting various sources and by basing decisions on the most reliable of those sources.[93] Similarly, in the context of AI-generated information, empirical evidence has shown that information consumers are very critical of AI-generated content.[94] We can rest assured that well-informed users are unlikely to unquestioningly accept chatbot output. Rather, the users will rely on chatbots as one source of information while simultaneously consulting other information sources—a cross-checking process that, in a competitive marketplace, should lead the users to the most trustworthy sources and the most accurate information.[95] In other words, discretion should be left to information consumers, enabling them to autonomously select information outlets according to their own preferences. This is also aligned with the spirit of the First Amendment, as the First Amendment not only protects the rights of speakers, but also the rights of listeners.[96] Admittedly, some consumers are not critical, in which case policymakers should bridge the wisdom gap by adopting transparency rules, AI literacy programs, and user empowerment schemes.

---

[89] *See* Simon et al., *supra* note 18, at 5 (arguing that the concerns over the spread of misinformation by generative AI are overblown).

[90] Here, I avoid using the wording "rational" because in the economic sense, "rational[ity]" entails the ability of "a cognitively intensive, calculating process of maximization in the self-interest." Vernon L. Smith, *Rational Choice: The Contrast Between Economics and Psychology*, 99 J. POL. ECON. 877, 878 (1991). However, this proves impossible in most cases when people make decisions in their daily lives. *See id.* at 880. Similarly, in the information market, the rationality requirement entails the capability of telling the normative good information and bad information, truth and falsity. However, this is impossible in most cases because, first, some information, especially opinions and ideas, unlike facts, cannot be proved false and thus inferior to other information. *See, e.g.*, Gertz v. Robert Welch, Inc., 418 U.S. 323, 340 (1974). And second, consumers also have inherent biases and cannot necessarily make rational choices. *See, e.g.*, Frederick Schauer, *Facts and the First Amendment*, 57 UCLA L. REV. 897, 909–10 n.64 (2010); Lyrissa Barnett Lidsky, *Nobody's Fools: The Rational Audience as First Amendment Ideal*, 2010 U. ILL. L. REV. 799, pt. III (2010). However, the fact that information consumers are not perfectly rational does *not* mean that they are not critical and diligent when processing information and that the market does not work effectively to check the incumbents. The effectiveness of market competition *never* relies on the *perfect rationality* of *all members* of society. In other words, it is sufficient to check the chatbot companies if some (not all) people are critical (not rational) when processing the output of the AI chatbots.

[91] *See, e.g.*, Sacha Altay & Alberto Acerbi, *People Believe Misinformation is a Threat Because They Assume Others Are Gullible*, NEW MEDIA & SOC'Y 1 (2023) (arguing that most people are not as gullible as the alarmists believe, and so the concerns over the prevalence and impact of misinformation are greatly exaggerated); Gordon Pennycook & David G. Rand, *Fighting Misinformation on Social Media Using Crowdsourced Judgments of News Source Quality*, 116 PROC. NAT'L ACAD. SCIS. 2521, 2521 (2019) (finding that "laypeople—on average—are quite good at distinguishing between lower- and higher-quality sources").

[92] *See* Sacha Altay et al., *Quantifying the "Infodemic": People Turned to Trustworthy News Outlets During the 2020 Coronavirus Pandemic*, 2 J. QUANTITATIVE DESCRIPTION: DIGIT. MEDIA 1 (2022).

[93] *See* María Celeste Wagner & Pablo J. Boczkowski, *The Reception of Fake News: The Interpretations and Practices That Shape the Consumption of Perceived Misinformation*, 7 DIGIT. JOURNALISM 870, 876–80 (2019) (introducing various strategies that online users adopt to check the information quality).

One potential objection here is that non-AI information outlets are not, or will soon not be, effective competitors of chatbot output. This postulation rests on the assumption that information-seeking users will grow over-reliant on chatbots if their output tends to be speedy and of high quality. A corollary of this argument is that this overreliance can be additionally harmful insofar as it promotes neglectful indolence in users, who may stop verifying the quality of AI-generated information even though the market may be competitive. The result would be a firm but unfounded belief in, or satisfaction with, patently erroneous or low-quality information.[97] This argument overlooks the nature of market competition, where consumers become loyal to a specific brand owing to its high quality. From a market perspective, this outcome is not problematic. It is even unproblematic if the quality of the revered brand decreases, because as the welfare of the brand's devoted customer base declines, the critical consumers in that base can and will switch to superior brands. Unlike governmental regulation, marketplaces chiefly care about competitive relationships between market players (*e.g.*, information outlets) and very little, if at all, about the quality of a specific commodity.[98]

---

[94] *See, e.g.*, Chiara Longoni et al., *News from Generative Artificial Intelligence Is Believed Less*, FAccT '22: PROC. OF THE 2022 ACM CONF. ON FAIRNESS, ACCOUNTABILITY, AND TRANSPARENCY 97, 97 (2022), https://dl.acm.org/doi/10.1145/3531146.3533077 [https://perma.cc/6LKW-84U4] (finding that "[p]eople were more likely to incorrectly rate news headlines written by AI (vs. a human) as inaccurate when they were actually true, and more likely to correctly rate them as inaccurate when they were indeed false").

[95] *See, e.g.*, Derek E. Bambauer & Mihai Surdeanu, *Authorbots*, 3 J. FREE SPEECH L. 375, 387 (2023) (arguing that "[f]ew take ChatGPT responses seriously at present, given its reputation as a hallucination engine, and that number falls further among observers who understand something about machine learning"); Simon et al., *supra* note 18, at 3–5 (arguing that the increase in quantity and quality of misinformation can hardly mislead the overall public).

[96] *See* Eugene Volokh et al., *Freedom of Speech and AI Output*, 3 J. FREE SPEECH L. 651, 651 (2023) ("AI programs' speech should be protected because of the rights of their users—both the users' rights to listen and their rights to speak."); Cass R. Sunstein, *Cass R. Sunstein: "Does Artificial Intelligence Have the Right to Freedom of Speech?"*, NETWORK L. REV. (Feb. 28, 2024), https://www.networklawreview.org/sunstein-artificial-intelligence/ [https://perma.cc/7ZXZ-Y9DS] ("But even if AI lacks free speech rights, the human beings who interact with generative AI, or with AI more broadly, have free speech rights, insofar as they are acting as speakers, and also insofar as they are acting as listeners, readers, or viewers.").

[97] *See* Risks of LLMs, *supra* note 2, at 23 ("[A] LM that gives factually correct predictions 99% of the time, may pose a greater hazard than one that gives correct predictions 50% of the time, as it is more likely that people would develop heavy reliance on the former LM leading to more serious consequences when its predictions are mistaken.").

A major concern can be that chatbot output may have silencing effects on non-AI sources of information and may thus harm the information market in the long run. For instance, a jurisdiction that allows ChatGPT to access, copy, and present the full version of online articles by the New York Times can deprive the newspaper of huge revenues because ChatGPT users, rather than pay the company a subscription fee, can access the articles at no additional cost from the chatbot. Such free-riding is anticompetitive and will inevitably disincentivize the copyright holders, in turn harming the overall information market and, ultimately, the money-saving information consumers themselves. Similarly, if ChatGPT is allowed to cater to some people's need to pry into other people's privacy by disclosing others' data, people would be inclined to keep information private and entirely inaccessible to the public. This would lead to a significant loss in the quantity and quality of data in the long run. As these harms to the information market are mostly indirect and can only be perceived over the long term, consumers, some of whom are short-sighted,[99] are inclined to overlook the harms. These scenarios are market failures, and I acknowledge them and support regulatory strategies that proportionately mitigate the aforementioned harms.[100]

## IV. RE-EVALUATIONS & SUGGESTIONS

Having reviewed how powerful the information market is at mitigating content risks, this Part examines some regulatory designs and generates policy suggestions. Drawing upon the inter-informational competitive relationship illustrated above, Section A re-evaluates the most popular regulatory proposals across the world. Then, Section B proposes three suggestions regarding: (a) concentration tendencies and potential anticompetitive practices in the internal market; (b) privacy disclosure; and (c) copyright-infringement risks in the external market.

---

[98] *See* Stacey L. Dogan & Mark A. Lemley, *Antitrust Law and Regulatory Gaming*, 87 TEX. L. REV. 685, 697 (2009) (explaining the differences between antitrust law and regulatory approaches).

[99] Economists have long recognized that consumers exhibit present bias, pursuing immediate satisfaction but overlooking long-term costs. *See, e.g.*, Ted O'Donoghue & Matthew Rabin, *Addiction and Present-Biased Preferences* (UC Berkeley, Paper No. E02-312, 2002). Such present-biased behaviors have been found in various fields and influence public policies. *See, e.g.*, David Bradford et al., *Time Preferences and Consumer Behavior*, 55 J. RISK & UNCERTAINTY 119 (2017).

[100] *See* discussion *infra* Section IV.B.2.

### A. *Re-evaluations of Existing Regulatory Proposals*

Policymakers and researchers have designed various regulatory proposals to mitigate the recognized risks of generative AI. While some challenge the effectiveness of these proposals,[101] a more fundamental question is: even if effective, are such regulations necessary? The key consideration here is whether the proposed regulation can achieve outcomes that market competition cannot. One imperative, yet poorly understood point should be made here: not all regulatory approaches *capable* of controlling target risks are *necessary*. Although some regulatory approaches promise rapid and ostensibly convenient solutions, heavy-handed interventions can have long-term, substantial negative fallout that weakens an entire industry.[102] In comparison, sufficient market competition may be able to accomplish the same end or an even better end. By this yardstick, some regulatory proposals are unnecessary. Proposals that survive the necessity test should also be proportionate: designed to invigorate information market competition and address anticompetitive issues that a free market cannot resolve on its own.

### 1. *Mandatory Content-Based Prohibition*

In the context of generative AI, mandatory prohibition entails that the government both ban problematic chatbot output and prescribe penalties for violators of the ban. For example, China has demonstrated a fervent commitment to assertive regulatory measures targeting information content. Article 4 of China's Generative AI Measures adopts a zero-risk approach by providing an exhaustive list of prohibited content.[103] In instances of non-compliance, relevant authorities may, as stipulated in Article 21, issue a warning or circular of reprimand, mandate corrections within a specified timeframe, and even institute an order to suspend service provisions.[104] Previously, the European Parliament version of the European Union AI Act exhibited traits similar to those of the Chinese measures.[105]

---

[101] *See, e.g.*, Neel Guha et al., *The AI Regulatory Alignment Problem*, HAI POL'Y BRIEFS (Nov. 2023), https://hai.stanford.edu/sites/default/files/2023-11/AI-Regulatory-Alignment.pdf [https://perma.cc/DMK5-L5D2] (arguing that "[s]ome proposals may fail to address the problems they set out to solve due to technical or institutional constraints").

[102] *See supra* notes 39–40 and accompanying text (the vicious cycle of the overreliance on direct regulation); *see also* Neel Guha et al., *AI Regulation Has Its Own Alignment Problem: The Technical and Institutional Feasibility of Disclosure, Registration, Licensing, and Auditing*, 92 GEO. WASH. L. REV. (forthcoming 2024) (manuscript at 6) (on file with SSRN), https://papers.ssrn.com/sol3/papers.cfm?abstract_id=4634443 [https://perma.cc/DN4E-8T96] (arguing that "some regulation proposals could—even if potentially useful in advancing legitimate public objectives—function to advantage powerful incumbents in AI and reduce competition, thus stymieing innovation and concentrating AI's benefits").

[103] *See* Chinese Generative AI Measures, *supra* note 9, art. 4(1)–(4).

Internal and external market competition already creates sufficient incentives for generative AI service providers to derisk their systems as much as possible in a bid to improve service quality and meet public expectations.[106] Although market competition can by no means ensure completely lawful content generation, neither can any regulations. Experts on generative AI have already generally acknowledged that the whole process of AI value chains entails inherent content risks.[107] Neither prohibitive regulation of unlawful output nor punitive treatment of violators can guarantee perfect observance of the regulation.[108]

Setting regulatory guardrails to act as bottom lines is sometimes necessary to protect the general public from being harmed and manipulated. This is especially important when corporate commercial interests do not align with public interests, which may lead to market failures and a race to the bottom.[109] However, in the context of chatbot content governance, AI chatbot companies under sufficient market forces must align their operations with public expectations. This means that imposing stringent content-based rules is unnecessary because AI chatbot companies have inherent incentives to compete for higher market share by addressing harmful content, discrimination, and misinformation risks continuously, automatically, and voluntarily. That said, policymakers might still consider prohibiting certain harmful content, such as hate speech and deepfakes, that pose significant risks to the broader information ecosystem.[110]

---

[104] *Id.* at art. 21.

[105] *See* Amendments Adopted by the European Parliament on 14 June 2023 on the Proposal for a Regulation of the European Parliament and of the Council on Laying Down Harmonised Rules on Artificial Intelligence (Artificial Intelligence Act) and Amending Certain Union Legislative Acts, EUR. PARL. DOC. (COM(2021)0206 – C9-0146/2021 – 2021/0106(COD)) §28b(4)(b) (2023), https://www.europarl.europa.eu/doceo/document/TA-9-2023-0236_EN.pdf [https://perma.cc/B8UT-DAPP].

[106] *See* discussion *supra* Section III.B.

[107] AI value chain is "the organisational process through which an individual AI system is developed and then put into use (or deployed)," involving various parties, including developers, deployers, and users. *See, e.g.*, Alex C. Engler & Andrea Renda, *Reconciling the AI Value Chain with the EU's Artificial Intelligence Act* 2 (CEPS 5, Sept. 2022), https://www.ceps.eu/ceps-publications/reconciling-the-ai-value-chain-with-the-eus-artificial-intelligence-act/ [https://perma.cc/B3SJ-R67M]; Tonja Jacobi & Matthew Sag, *We Are the AI Problem*, 74 EMORY L.J. ONLINE 1 (2024).

[108] For a detailed explanation of why the mandatory prohibitions based on the zero-risk goals are fundamentally flawed, see Jiawei Zhang, *ChatGPT as the Marketplace of Ideas: Should Truth-Seeking Be the Goal of AI Content Governance?*, 35 STAN. L. & POL'Y REV. ONLINE 11 (2024).

[109] *See, e.g.*, MAURICE E. STUCKE & ARIEL EZRACHI, COMPETITION OVERDOSE: HOW FREE MARKET MYTHOLOGY TRANSFORMED US FROM CITIZEN KINGS TO MARKET SERVANTS, ch. 1 (2020).

Regulators must remain cognizant of the virtues of market mechanisms and the shortcomings of content-based rules. Contrary to the assumptions of some policymakers, onerous direct regulations may have many side effects. First, unnecessary regulations raise market entry barriers and thus scare off would-be market entrants. This outcome should not be surprising, as startups can hardly bear the high costs that accompany both regulatory compliance and regulatory punishment.[111] Even worse, such restrictions can also have chilling effects on existing generative AI systems. To comply with the rules, established chatbot providers will over-moderate their models to make them needlessly prudential. Mandating a zero-risk approach or imposing a rigid truth-telling obligation could transform a dynamic tool like ChatGPT into a stifled and overly cautious "SorryGPT"—users will receive tons of apologies without any meaningful content. This reactive approach ultimately burdens consumers, diminishing their access to desired information and eroding their overall surplus.

### 2. Licensure

Another noteworthy regulatory proposal, this one representing a variant of mandatory prohibitions, is licensure.[112] It requires that generative AI service providers, before they even enter the market, acquire a government-issued license by demonstrating a sufficient level of fairness and accuracy.[113] A provider's failure to demonstrate these capabilities would lead the government to adopt a presumption of unlawfulness with regard to the provider's potential, as-yet-unrealized provision of services.[114] In this sense, licensure, as an *ex-ante* scheme constraining generative AI services, and mandatory prohibition proposals are essentially two sides of the same coin.

---

[110] *See, e.g.*, Jiawei Zhang, *Informational Public Interest*, 26 N.C. J.L. & TECH. 73, 116–17 (2024).

[111] *See, e.g.*, Thibault Schrepel, *Decoding the AI Act: A Critical Guide for Competition Experts* 11–18 (ALTI Working Paper Series 2024) (arguing that the EU AI Act may distort market competition because it is not technology neutral in nature, distributes the regulatory burden unevenly, and significantly raises the threshold of market entry).

[112] *See* Gianclaudio Malgieri & Frank Pasquale, *Licensing High-Risk Artificial Intelligence: Toward Ex Ante Justification for a Disruptive Technology*, 52 COMPUT. L. & SEC. REV. 105899 (2024). Interestingly, Sam Altman urges the establishment of the licensure system and punishment of wrongful conduct. *See* Sindhu Sundar, *Sam Altman Says a Government Agency Should License AI Companies — and Punish Them If They Do Wrong*, BUS. INSIDER (May 16, 2023), https://www.businessinsider.com/sam-altman-openai-chatgpt-government-agency-should-license-ai-work-2023-5 [https://perma.cc/74RQ-QVD5].

[113] *See* Malgieri & Pasquale, *supra* note 112, at 1.

[114] *Id.* ("Under a licensing system, products, services, and activities are unlawful until the entity seeking to develop, sell, or use them has proven otherwise.").

The reasons why mandatory prohibition is likely unnecessary are the same reasons why licensure is likely unnecessary. However, a licensure regime can be even more problematic than mandatory prohibition because the licensure standard for pre-market approval is even more rigid. More than mandatory prohibition, it will erect an insurmountable market-entry barrier for many would-be market entrants and will further entrench incumbents, a scenario that will ultimately undermine internal-market competition and harshen the duties on those incumbents.[115]

### 3.   *Curation of Datasets*

Dataset curation rules require generative AI service providers to eliminate some content risks by purging potentially problematic materials from upstream datasets. As a root-level regulatory tool, dataset curation rules are supposed to address some content risks, such as discriminatory content,[116] privacy disclosure,[117] and copyright infringement.[118]

However, imposing data curation rules on chatbot companies in a bid to counter content risks is yet another unnecessary and even detrimental regulatory proposal. Take, for example, the application of data curation to copyright issues. The New York Times has recently sued OpenAI and Microsoft, calling for chatbot companies to destroy their datasets that contain materials copyrighted by the newspaper.[119] Undoubtedly, copyright holders hoping to maximize economic benefits have the right to file a lawsuit against generative AI service providers. However, AI experts have recognized that it is technically impossible for AI companies to filter out all copyrighted material from their datasets.[120] And, it is still highly uncertain whether LLMs can maintain adequate performance if they are allowed to access only public, non-copyrightable, and permissively licensed data.[121] OpenAI, in response to the rising tensions between it and copyright holders, has warned that a copyright crackdown will make its services "impossible."[122] OpenAI further explained that AI systems trained merely on public domain data "would not . . . meet the needs of today's citizens."[123]

---

[115] *See* Guha et al., *supra* note 102, at 52–53.

[116] *See* Philipp Hacker et al., *Regulating ChatGPT and Other Large Generative AI Models*, ACM CONF. ON FAIRNESS, ACCOUNTABILITY, & TRANSPARENCY 1112, 1119–20 (2023).

[117] *See* Amy Winograd, *Loose-Lipped Large Language Models Spill Your Secrets: The Privacy Implications of Large Language Models*, 36 HARV. J.L. & TECH. 615, 649–51 (2023) (The "publicly-intended data" indicates "data that is most likely to be intended for broad public consumption and use in a wide variety of contexts.").

[118] *See* White Paper: How the Pervasive Copying of Expressive Works to Train and Fuel Generative Artificial Intelligence Systems Is Copyright Infringement and Not a Fair Use, NEWS/MEDIA ALLIANCE (Oct. 31, 2023), https://www.newsmediaalliance.org/generative-ai-white-paper/ [https://perma.cc/H5ZP-MVTH].

[119] *See* Michael M. Grynbaum & Ryan Mac, *The Times Sues OpenAI and Microsoft Over A.I. Use of Copyrighted Work*, N.Y. TIMES (Dec. 27, 2023), https://www.nytimes.com/2023/12/27 /business/media/new-york-times-open-ai-microsoft-lawsuit.html [https://perma.cc/6HZP-LD5L].

[120] *See* Peter Henderson et al., *Foundation Models and Fair Use* (Mar. 29, 2023) (manuscript at 22) (unpublished manuscript) (on file with arXiv), https://arxiv.org/abs/2303.15715 [https://perma.cc/MS3V-2XPQ].

I fully agree with OpenAI on these points and argue that the Times' claims should be roundly rejected by judges and policymakers alike. Chatbot companies have inherent incentives to curate their datasets to improve their provided services and better cater to public needs.[124] Compulsory data filtering would be catastrophic for the whole industry, not only demotivating potential entrants but also stagnating incumbents' innovation. I concur that both free riding off others' intellectual creation and privacy disclosure are real concerns for a healthy, competitive, inter-informational environment.[125] However, rather than rely on data curation rules, policymakers should design alternative regulatory strategies that do not weaken but strengthen the beneficial powers of competitive markets.[126]

### 4. Transparency

Transparency rules have gained widespread acceptance as a regulatory approach to AI risks and are attractive given the essentially opaque nature of AI technology.[127] In the context of generative AI, transparency rules require that chatbot companies practice self-disclosure by informing users and public officials of relevant information about the scope of training data sources, the performance of the LLMs, and measures that companies take to handle existing and potential risks.[128] Embracing a pro-innovation approach, the UK government in March 2023 published its AI regulation guidelines in a white paper, endorsing a more light-touch and technology-neutral regulatory approach.[129] Specifically, the UK government is trying to harness transparency rules that enhance users' understanding of and trust in generative AI products.[130]

---

[121] *Id.*

[122] *See* James Titcomb & James Warrington, *OpenAI Warns Copyright Crackdown Could Doom ChatGPT*, THE TELEGRAPH (Jan. 7, 2024), https://www.telegraph.co.uk/business/2024/01/07/openai-warns-copyright-crackdown-could-doom-chatgpt/ [https://perma.cc/LP3L-BGMU].

[123] *Id.*

[124] In fact, chatbot companies, such as OpenAI, have voluntarily taken actions to moderate the problematic data in their datasets to reduce hallucination and privacy disclosure. *See* OpenAI's Report, *supra* note 2, at 46, 53.

[125] *See* discussion *supra* Section III.B.2.

[126] *See* discussion *infra* Section IV.B.2.

[127] *See generally* Warren J. von Eschenbach, *Transparency and the Black Box Problem: Why We Do Not Trust AI*, 34 PHILOS. TECH. 1607 (2021); *see also* Bruno Lepri et al., *Fair, Transparent, and Accountable Algorithmic Decision-Making Processes*, 31 PHILOS. TECHNOL. 611, 613 (2018).

[128] For a general proposal to govern AI risks, see, for example, Sonia K. Katyal, *Private Accountability in the Age of Artificial Intelligence*, 66 UCLA L. REV. 54, 111–17 (2019); Hacker et al., *supra* note 116, at 1119; for a proposal to adopt transparency rules to control specific generative AI risks, such as privacy, see, for example, Winograd, *supra* note 117, at 652–54 (arguing that "developers should be required to disclose the sources of training data, the measures taken to ensure data collection and processing conforms to applicable law, and the privacy-preserving and safety measures employed in training and implementation to ensure responsible development.").

Transparency is one of the rules that passes the necessity test. As the "best of disinfectants,"[131] transparency has proved effective at mitigating risks and invigorating competition in an array of market types. In traditional physical markets, for example, transparency rules require food producers to inform consumers of a food product's ingredients, nutrients, health hazards, and other information that helps consumers make informed comparisons and decisions in the marketplace.[132]

In digital markets, the U.S. government has long embraced the transparency rule as an ideal candidate for a light-touch regulatory framework. One noteworthy example is the 2018 Internet Order, in which the FCC, to regulate the broadband internet, shifted away from a network neutrality framework and toward a market-based one, thereby abandoning the no-blocking, no-throttling, and no-paid-prioritization rules while retaining the transparency rule.[133] The FCC believed that transparency rules help the market achieve self-regulation.[134] When ISPs appropriately disclose relevant information, regulators can dynamically grasp and monitor market conditions, other market entities can gauge highly topical matters such as fairness, users can make better choices confidently, and the ISPs themselves can promptly self-correct in instances of misconduct.[135]

The chatbot market is no exception to the above trends for the transparency rule. Researchers at Stanford's Institute for Human-Centered Artificial Intelligence have comprehensively evaluated the proposed disclosure requirements and yielded four central findings: (1) disclosure rules are relatively easier for governments to implement;[136] (2) good disclosure rules are understandable and actionable for consumers and are verifiable for experts including regulatory agencies;[137] (3) effective disclosure rules can be administratively costly;[138] and (4) disclosure rules should therefore be as proportionate as possible.[139] Here, some additional remarks are attached to these four findings.

---

[129] *See* UK DEPARTMENT FOR SCIENCE, INNOVATION AND TECHNOLOGY, AI REGULATION: A PRO-INNOVATION APPROACH, 2023, Cp. 815 (UK), https://www.gov.uk/government/publications/ai-regulation-a-pro-innovation-approach/white-paper [perma.cc/U2PE-8LL2] [hereinafter WHITE PAPER or UK AI REGULATION WHITE PAPER] (In the context of LLMs, the White Paper believes that "it would be premature to take specific regulatory action in response to foundation models including LLMs . . . [which] would risk stifling innovation, preventing AI adoption, and distorting the UK's thriving AI ecosystem.").

[130] *Id.* at 58.

[131] LOUIS D. BRANDEIS, OTHER PEOPLE'S MONEY AND HOW THE BANKERS USE IT 92 (1914) ("Sunlight is said to be the best of disinfectants.").

[132] In fact, the AI transparency rules have been compared to the "nutrition facts labels" of food products. *See* Sara Gerke, *"Nutrition Facts Labels" for Artificial Intelligence/Machine Learning-Based Medical Devices—The Urgent Need for Labeling Standards*, 91 GEO. WASH. L. REV. 79 (2023).

[133] *See* 2018 Internet Order, *supra* note 44, at 124–25.

[134] *See id.* at 125.

[135] Guha et al., *supra* note 102, at 23.

First, I agree that disclosure rules are easier for governments to implement than other rules. This relative ease is rooted primarily in the fact that disclosure rules, unlike most other rules, shift much of the information-processing burden from government enforcement agencies to regulated entities.[140] Second, I also agree that the effectiveness of disclosure rules is subject to how understandable and actionable the rules are for the consumers and how verifiable a company's adherence to the rules is for experts (*e.g.*, government supervisory agencies).[141] In a market sense, understandability is a prerequisite for consumers seeking to compare various AI products or services with one another; likewise, the comparisons must enable consumers to take informed action: stay with the present offering or switch to an alternative. As for verifiability, it ensures the accuracy of the information disclosed and safeguards a healthy competitive environment.

Third, it is true that policymakers will invariably bear some administrative burden to design well-defined and proportionate disclosure rules. However, any new rules entail unavoidable administrative costs, and this fact should never excuse lassitude in policymaking. Therefore, and lastly, policymakers should proportionately tailor disclosure rules to prevent the entirely avoidable chilling effects that overbroad disclosure rules can have on private entities, including generative AI companies.[142] For example, policymakers should be aware of the possible tension that can exist between disclosure rules and trade secret protection, as enlarged transparency rules may unreasonably mandate the disclosure of information that companies have good reason to withhold from the public.[143]

---

[136] *Id.* at 24 (noting that "[s]everal characteristics make disclosure requirements easier to implement than other regulatory interventions").

[137] *Id.* at 26–27 (noting that "disclosures are most effective when they meet three criteria: understandability, actionability, and verifiability").

[138] *Id.* at 21–22, 25–26.

[139] *Id.* at 29–30.

[140] *Id.* at 24.

[141] *Id.* at 26–29.

[142] *Id.* at 29–30.

[143] *See* Katyal, *supra* note 128, at 117–21.

### 5.  *Traceability*

Given the increasingly ambiguous boundary between machine output and human speech, traceability rules have been adopted to prevent bad actors from misusing AI-generated content in ways that distort our understanding.[144] There are two main pillars for public traceability of chatbot-generated content. The first pillar is that generative AI service providers must make their generated content technically detectable by developing and employing technological methods for such tasks, such as watermarking.[145] The second pillar is that users of generative AI products must disclose their uses to relevant parties.[146] Section 4.5 of the U.S. AI Order requires that the Secretary of Commerce consider and evaluate the following three dimensions: (1) "authenticating content and tracking its provenance," (2) "detecting synthetic content," and (3) labeling it with, for example, watermarks.[147]

The two pillars of the traceability proposal are necessary but should be subject to the proportionality test. The first pillar is necessary because some sensitive AI-generated information, especially when in audio and visual formats, has great potential to confuse and deceive the public. To make matters worse, the self-regulatory mechanisms of the information market do not effectively counter this threat. For example, malicious pornographers can morph celebrities' heads into sexually explicit deepfakes.[148] AI can also be used to generate fake audio and visual output that can, for example, deceive voters in the run-up to an election.[149] In all of these cases, inter-informational competition is doomed to fail if traceability is not possible, because, in the absence of traceability, people will reasonably assume that the audio and visuals are conveying "facts." The only other recourse would be for people to doubt the authenticity of all audio and visual evidence—an outcome that is hugely destructive in its own right.

---

[144] *See, e.g.*, Alexei Grinbaum & Laurynas Adomaitis, *The Ethical Need for Watermarks in Machine-Generated Language* (Sept. 7, 2022) (unpublished manuscript) (on file with arXiv), https://arxiv.org/pdf/2209.03118 [https://perma.cc/2DUJ-YSRT].

[145] *See, e.g.*, Hacker et al., *supra* note 116, at 1119.

[146] *Id.*

[147] *See* Exec. Order No. 14,110, 88 Fed. Reg. 75,191, § 4.5, 75, 202–03 (Nov. 1, 2023) (to be codified at 3 C.F.R.), https://www.govinfo.gov/content/pkg/FR-2023-11-01/pdf/2023-24283.pdf [https://perma.cc/AVU5-DQVU] [hereinafter Order or U.S. AI Order].

[148] *See, e.g.*, Matt Burgess, *Deepfake Porn Is Out of Control*, WIRED (Oct. 16, 2023), https://www.wired.com/story/deepfake-porn-is-out-of-control/ [https://perma.cc/7KL2-NSNQ].

[149] For example, an audio recording surfaced on Facebook purportedly capturing Michal Šimečka, leader of the Progressive Slovakia party, and Monika Tódová from Denník N discussing election manipulation, including the purchase of votes from the marginalized Roma minority just two days before the election began. *See* Morgan Meaker, *Slovakia's Election Deepfakes Show AI Is a Danger to Democracy*, WIRED (Oct. 3, 2023), https://www.wired.com/story/slovakias-election-deepfakes-show-ai-is-a-danger-to-democracy/ [https://perma.cc/SVJ5-AES5].

Generally, statements of facts are considered falsifiable.[150] But in the age of generative AI, the absence of traceability for AI-generated audio and visuals explodes the possibility of falsifiability in many instances. Consequently, people will have no clue as to whether a proposed piece of evidence is true or false: is it an authentic "capture" of reality or an AI-generated fiction? Disinformation will flood the public square, profoundly and destructively shaping the information marketplace. However, with the aid of traceability rules, the public can detect AI-generated content for what it is and can thus distinguish it from genuine content. Traceability offers the promise of deeply undercutting the competitiveness of deepfakes in the information marketplace.

However, despite its usefulness, the traceability rule should be limited to those contexts that actually threaten competition in the information marketplace. There should be no government requirement that AI service providers watermark their textual output.[151] The underlying reason for this limit is that text-based disinformation is not nearly as disruptive to truth-finding as audio and visual disinformation is. Although we can imagine cases where chatbot users generate, for example, a pornographic text featuring a certain celebrity or a news article regarding an election bribe that never took place, these texts are subject to effective verification. Thus, the information marketplace for purely textual chatbot output is quite competitive. Additionally, context-based traceability lacks a great deal of feasibility.[152] For example, research found that the detection tool can only correctly identify 26% of AI-written text and produces false positives 9% of the time.[153]

According to the second pillar, users of generative AI should disclose their use in specific fields. I support this rule because it is entirely technology-neutral and poses no harm to competition in the information marketplace.[154] Any technology has its evil side. Consider knives; they are potentially deadly, but we never limit their sharpness. Generative AI is no exception. Therefore, apart from some special fields, such as healthcare and legal services, that need legislation governing disclosure requirements, industry's self-generated guidelines should be sufficient for private-sector entities seeking to establish when and how users should disclose their use of generative AI.

---

[150] *See* Mann v. Abel, 10 N.Y.3d 271, 276 (2008) (ruling that "[e]xpressions of opinion, as opposed to assertions of fact, are deemed privileged and, no matter how offensive, cannot be the subject of an action for defamation").

[151] *See, e.g.*, Peter Henderson, *Should the United States or the European Union Follow China's Lead and Require Watermarks for Generative AI?*, GEO. J. FOR INT'L AFFS. (May 24, 2023), https://gjia.georgetown.edu/2023/05/24/should-the-united-states-or-the-european-union-follow-chinas-lead-and-require-watermarks-for-generative-ai/ [https://perma.cc/XL8P-92WM].

[152] *Id.*

[153] *See* Fiona Jackson, *OpenAI Quietly Bins Its AI Detection Tool — Because It Doesn't Work*, TECHHQ (July 27, 2023), https://techhq.com/2023/07/ai-classifier-tool-cancelled-ineffective/ [https://perma.cc/KHT6-MHYJ].

### 6. *Notice-and-Respond Mechanism*

Another highly popular regulatory proposal is the notice-and-respond mechanism. The rule behind it stipulates that generative AI service providers should actively mitigate content risks by expeditiously responding to users' reports of problematic output. This mechanism bears a resemblance to the well-known "safe harbor" rules in copyright law.[155] In contrast to the dataset curation rule, which acts as a root-level cleansing approach, the notice-and-respond mechanism is decentralized and can control risks at the user's end.[156] Some EU researchers have suggested that the EU should expand its Digital Services Act (DSA),[157] especially the "notice and action mechanism"[158] and the appointment of "trusted flaggers." [159] Some U.S. researchers have similarly endorsed this bottom-up approach to controlling misinformation,[160] privacy,[161] and copyright risks.[162]

Although the proposal to expand the notice-and-respond mechanism sounds impressive, its necessity is still questionable. The notice-and-respond mechanism requires chatbot companies to carry out two basic tasks: (1) establish a mechanism through which the users could give the providers rapid and sufficiently detailed feedback on generated content; and (2) respond promptly and satisfactorily to users' feedback. A failure on the part of the providers to carry out either task would subject them to severe penalties from enforcers, or to the threat of civil cases from users seeking to prove that the providers are liable for "actual malice" or "negligence."[163]

---

[154] UK's policymakers have raised a similar idea that a clear, proportionate approach to regulation entails regulatory measures focusing on the use of AI, rather than the technology itself. *See* UK AI REGULATION WHITE PAPER, *supra* note 129, at 5.

[155] *See* 17 U.S.C. § 512(c)(1)(C) (Service providers should, "upon notification of claimed infringement as described in paragraph (3), respond[] expeditiously to remove, or disable access to, the material that is claimed to be infringing or to be the subject of infringing activity").

[156] *See* Hacker et al., *supra* note 116, at 1120.

[157] As the applicable scope of the DSA rules is confined to the "intermediary service," including "mere conduit service," "cashing service," and "hosting service," the LLM systems are not subject to DSA obligations. *See id.* at 1118.

[158] *See* Regulation (EU) 2022/2065 of the European Parliament and of the Council of 19 October 2022 on a Single Market For Digital Services and amending Directive 2000/31/EC (Digital Services Act), 2022 O.J. (L 277) 50 [hereinafter DSA].

[159] *Id.* at 56.

[160] *See, e.g.*, Eugene Volokh, *Large Libel Models? Liability for AI Output*, 3 J. FREE SPEECH L. 489, 518–20 (2023) (proposing to use "notice-and-blocking" regime to mitigate the risks of false and defamatory assertions generated by AI systems).

[161] *See, e.g.*, Jon M. Garon, *An AI's Picture Paints a Thousand Lies: Designating Responsibility for Visual Libel*, 3 J. FREE SPEECH L. 425, 445–52 (2023) (proposing "notice-and-takedown" rules to mitigate libel and privacy risks).

[162] *See, e.g.*, Henderson et al., *supra* note 120, at 26 (proposing to extend DMCA safe harbor rules to shield generative AI output from copyright infringement liabilities).

Some researchers have argued that a wealth of knowledge acquired from over two decades of experience with safe harbor rules in the realm of copyright law can be applied, with good effect, to the realm of generative AI.[164] However, the notice-and-takedown regime in copyright law would likely experience "transplant rejection" in the context of generative AI. The necessity of notice-and-takedown in copyright law is contingent on at least the following conditions: First, alleged copyright infringement occurs in the public sphere, meaning that a notice-and-takedown mechanism can, at least in theory, effectively stop the dissemination of copyright-infringing works. However, the use of chatbots typically occurs in a private, one-to-one context. And although chatbots generate problematic content, such as misleading or false information, there might be no harm to a specific user if that user is critical enough.[165] The problematic content will cause no substantial social harm as long as the user does not disseminate it to the general public.[166]

The second condition justifying notice-and-takedown mechanisms in copyright law is material clarity. Copyright infringement involves relatively objective and concrete evidence, so that copyright holders can issue a takedown notification to a copyright violator, requesting that it take action based on detailed, verifiable facts.[167] However, in the context of chatbots, the obvious lack of basic consensus regarding what qualifies as problematic information makes a valid notification from lay users impractical. Thus, the volume of hopelessly unclear, highly subjective notifications would burden AI companies with prohibitively high administrative costs that do not arise in the realm of copyright law.[168]

---

[163] *See, e.g.*, Volokh, *supra* note 160, at 514–21, 531.

[164] *See id.* at 519.

[165] *See supra* notes 90–94 and accompanying text.

[166] Therefore, I argue that the point of regulation should not be on technologies and AI companies, but on users. *See* discussion *supra* Section IV.A.4.

[167] *See* 17 U.S.C. § 512(c)(3)(A).

[168] In fact, the notice-and-takedown regime of copyright law has long been criticized as ineffective and inefficient, just like the whack-a-mole game. The internet service providers have to spare many resources that could be otherwise more productive in handling numerous notifications. *See, e.g.*, U.S. COPYRIGHT OFF., SECTION 512 OF TITLE 17: A REPORT OF THE REGISTER OF COPYRIGHTS 32, 76 (2020).

The third notice-and-takedown condition in copyright law is its feasibility and affordability of the take-down process. Some advocates of copyright safe harbor argue that "[t]he genius of the DMCA is that it lets technology startups comply with the law without hiring a platoon of copyright lawyers."[169] However, this ease would not be transferable to chatbots. Chatbot users' requests, comments, and complaints about problematic information would force AI researchers to spend an inordinate amount of time curating their datasets and training LLMs to produce answers that still would strike many users as unethical or inaccurate.[170] Therefore, the whole costly process would simply yield more dissatisfaction and uncertainty. In short, the notice-and-respond rule, when applied to chatbot companies, would not be a safe harbor, but dangerous waters.

My point is not that chatbot companies should steer clear of notice-and-respond mechanisms. These mechanisms should not be compulsory. Chatbot companies are already under market pressure to adopt such feedback mechanisms, which, to this extent, can help the companies improve the quality of their output with reasonable efficacy. To date, chatbot companies have voluntarily implemented such a mechanism.[171]

### 7. Audits

Proponents of regulations have widely argued that governments should subject generative AI providers to auditing rules, which would necessitate the establishment of external neutral auditors and the formulation of a tailored set of auditing procedures.[172] These procedures would help to identify, assess, monitor, and manage both risks and risk-management measures linked to chatbot companies.[173] Many jurisdictions have considered implementing auditing rules to exert greater control over the performance of generative AI service providers.[174]

---

[169] *See, e.g.*, Chris Sprigman & Mark Lemley, *Op-Ed: Why Notice-and-Takedown Is a Bit of Copyright Law Worth Saving*, L.A. TIMES (June 21, 2016), https://www.latimes.com/opinion/op-ed/la-oe-sprigman-lemley-notice-and-takedown-dmca-20160621-snap-story.html [https://perma.cc/C86W-UKNF].

[170] *See* Henderson et al., *supra* note 120, at 19 (explaining the challenges and uncertainties surrounding the application of DMCA takedowns in the context of generative models).

[171] *See, e.g.*, *supra* discussion and accompanying note 86.

[172] *See generally* Jakob Mökander et al., *Auditing Large Language Models: A Three-Layered Approach*, AI ETHICS (2023); Jakob Mökander & Luciano Floridi, *Ethics-Based Auditing to Develop Trustworthy AI*, 31 Minds & Machines 323 (2021).

[173] *See, e.g.*, Hacker et al., *supra* note 116, at 1120 (arguing for an expansion of the DSA rules (i.e., Art. 34–35) to establish comprehensive auditing rules to manage risks of LLM systems); *see also* Winograd, *supra* note 117, at 654–55 (suggesting the establishment of an oversight body to exercise comprehensive auditing work to ensure compliance and bolster the goal of privacy protection).

Although the regulatory design of a typical AI audit involves many unresolved problems and intractably difficult steps,[175] AI auditing is a necessary regulatory endeavor. First, auditing regulations are ancillary to established regulations that are directly related to chatbot output. At least on the surface, the objective of AI auditing is simple: to examine whether an AI company has taken necessary and sufficient measures to comply with the established laws.[176] This means that, in a light-touch and market-based regulatory environment, AI auditing will not be overly burdensome, as generative AI service providers that comply with mandatory auditing do not need to shoulder heavy responsibilities.

Under a light-touch regulatory auditing regime, neither an independent third party nor a government regulator would need to examine whether a chatbot's output is accurate, non-biased, ethical, and lawful because the light-touch regime would not have such requirements; instead, what would be in the purview of the audits is, for example, whether generative AI service providers have followed the transparency rules regarding understandability and accuracy.[177] In other words, as a non-substantive and indirect regulatory approach, AI audits would impose costs that are directly proportionate to the costs of direct regulation. As long as the direct conduct-based duties that the government imposes on AI companies remain proportionate, the auditing rules will remain proportionate too. In this sense, many concerns regarding the cost and feasibility of AI audits are unwarranted.[178]

---

[174] *See, e.g.*, U.S. AI Order, *supra* note 147, at 75196 ("[T]o help ensure the development of safe, secure, and trustworthy AI systems, [authorities] shall . . . (i) [e]stablish guidelines and best practices, . . . including: . . . (C) launching an initiative to create guidance and benchmarks for evaluating and auditing AI capabilities, with a focus on capabilities through which AI could cause harm, such as in the areas of cybersecurity and biosecurity."); UK AI REGULATION WHITE PAPER, *supra* note 129, at 64 ("To assure AI systems effectively, we need a toolbox of assurance techniques to measure, evaluate and communicate the trustworthiness of AI systems across the development and deployment life cycle. These techniques include impact assessment, audit, and performance testing along with formal verification methods."); Chinese Generative AI Measures, *supra* note 9, at art.19 ("The relevant authorities shall conduct supervisory inspections of generative AI services based on their respective functions and responsibilities . . .).

[175] *See* Evan Selinger et al., *AI Audits: Who, When, How . . . Or Even If?* (Sept. 11, 2023) (unpublished manuscript) (on file with SSRN), https://ssrn.com/abstract=4568208 [https://perma.cc/962Z-8QSL]; *see also* Guha et al., *supra* note 102, at 55–69.

[176] *See* Selinger et al., *supra* note 175, at 10.

[177] *See supra* discussion and accompanying note 141.

[178] For these concerns, see, for example, Guha et al., *supra* note 102, at 58–69.

144

### 8.  Liability

Rather than call for the design and implementation of new regulatory rules, some researchers have sought to address the risks of AI output by turning to existing rules. One proposal is to subject chatbot companies to private law obligations. In this way, private plaintiffs can have grounds to claim damages caused by chatbot output. Although a private plaintiff must enforce private law obligations, the deterrence acts similar to government-enforced regulations.[179]

Among various proposals, one would involve applying traditional product liability law to chatbot output.[180] This proposal is considered suitable for determining fault in cases involving emerging technologies.[181] When, for example, the design of a chatbot product has inherent defects that can harm consumers, a plaintiff can claim damages based on product liability.[182] Additionally, it has been argued that a finding of liability for an AI company should rest on—and be consistent with—established cases concerning, among other things, search-engine results and mass-media products.[183] Some experts in the field have compared chatbot companies' duty of care to parents of mischievous children or to owners of unruly pets.[184] When chatbot companies "know[] or [have] reason to know" that their content is legally unacceptable, liability rules that have been established for traditional information media, such as bookstores, newsstands, and property owners, should be applicable to the chatbot companies.[185]

Market competition cannot offer a company immunity from common law rules designed to protect customers. However, differences between traditional physical markets and today's information markets lie primarily in the fact that in the context of the information market, the harms to reputation, privacy, and so on tend to be intangible. Here, two points are noteworthy regarding how courts should handle the intangible harms and potential liability of chatbot companies. First, judges should ascertain that real harm, rather than mere risk, exists in a present case. As I argued previously, inter-informational competition in internal and external markets can alleviate most risks associated with chatbot content.[186] For example, in the event where ChatGPT baselessly accused a U.S. law professor of sexual harassment,[187] does the output cause any harm to the professor? Here, ChatGPT "cited" a March 2018 article in the Washington Post as the source of the information.[188] However, the cited article does not exist at all. Skeptics can effortlessly verify or disprove that information via other competing information outlets. Most people would not take the information seriously—this is little more than a fool wielding a bullhorn and spouting nonsense. However, that does not mean that no one is responsible for the dissemination of such rumors. People who use chatbots and disseminate the generated content may have a duty to check the sources of the content, because publishers who either have "knowledge of [the content's] falsity or [are] in reckless disregard of the truth" are ineligible for First Amendment protection and may incur liabilities of defamation.[189]

---

[179] One prevalent and broad reading of "regulation" is to treat it "as an activity that restricts behaviour and prevents the occurrence of certain undesirable activities," which can also be read as "risk management." In this sense, private law obligations and liabilities can be seen as a kind of regulation. *See* ROBERT BALDWIN, MARTIN CAVE & MARTIN LODGE, UNDERSTANDING REGULATION: THEORY, STRATEGY, AND PRACTICE 3 (2nd ed. 2012); BRONWEN MORGAN & KAREN YEUNG, AN INTRODUCTION TO LAW AND REGULATION: TEXT AND MATERIALS 13 (2007) (indicating that the concepts "regulation" and "risk management" can be used interchangeably); *see also* Douglas A. Kysar, *Public Life of Private Life: Tort Law as a Risk Regulation Mechanism*, 9 EUR. J. RISK REGUL. 48, 51 (2018).

[180] *See* Nina Brown, *Bots Behaving Badly: A Products Liability Approach to Chatbot-Generated Defamation*, 3 J. FREE SPEECH L. 389 (2023).

[181] *See id.* at 392.

[182] *See id.* at 410–11 (showing examples where the plaintiff may have an argument that the design was defective).

[183] *See* Jane Bambauer, *Negligent AI Speech: Some Thoughts About Duty*, 3 J. FREE SPEECH L. 343, 362 (2023).

[184] *Id.*

[185] *See* Volokh, *supra* note 160, at 520–21.

Second, if chatbot output causes harm, whether a court can find the chatbot company liable should still be subject to a risk-utility test. The Hand formula,[190] or more precisely the marginal Hand formula,[191] should govern the test. This formula can help courts to determine whether the marginal cost of a precaution that a chatbot company can take will exceed the marginal benefits. Despite some uncertainties surrounding its application, this formula at least shows that the policy advocating for zero risk is impossible to meet and is thus unreasonable for chatbot companies.[192] For certain risks beyond the duty of care that is owed by chatbots, the main responsibility is on chatbot users: they should exercise individual prudence to mitigate the risks associated with using a chatbot. Chatbots generate content in response to users' prompts, just as McDonald's fulfills customer orders for hot coffee—there are attendant risks, but if the service providers properly warn the customers about the main risks, the residual risks should rest with the end users. After all, McDonald's cannot ensure that every customer safely sips a hot coffee purchased from a franchise.[193]

---

[186] *See* discussion *supra* Section III.B.

[187] *See* Verma & Oremus, *supra* note 20.

[188] *Id.*

[189] Time, Inc. v. Hill, 385 U.S. 374, 387–88 (1967).

[190] *See* United States v. Carroll Towing Co., 159 F.2d 169, 173 (2d Cir. 1947) ("[I]f the probability be called P; the injury, L; and the burden, B; liability depends upon whether B is less than L multiplied by P: i.e., whether B less than PL.").

[191] *See* COOTER ROBERT & THOMAS ULEN, LAW AND ECONOMICS 214 (6th ed. 2014) ("The marginal Hand rule states that the injurer is negligent if the marginal cost of his or her precaution is less than the resulting marginal benefit. Thus, the injurer is liable under the Hand rule when further precaution is cost-justified.").

[192] *See, e.g.*, Volokh, *supra* note 160, at 526.

Essentially, I am arguing that chatbot companies should provide users with conspicuous and adequate warnings of potential risks. These warnings can take the form of prompts stating, "The answer may contain misleading information" or "Dissemination of this information may put you at risk of legal liability." In addition, chatbot output should include links or notes identifying the sources of the output. These steps on the part of chatbot companies should accomplish two desirable outcomes: (1) reduce copyright infringement risks by softening the competitive relationship between the chatbot output and copyrighted materials,[194] and (2) help users track the sources of chatbot output and verify the accuracy of the output.

## B.   *Further Suggestions*

Despite the suggestions previously made, there are information-market risks that have not yet been addressed. Below, I will present some preliminary thoughts on three of them. First, internal market risks are the increasing market concentration tendency and the negative effects that the anticompetitive practices of incumbent AI companies may have on market competition. These risks, in turn, will reduce the companies' incentives to improve the quality of their chatbot output. The other two risks are privacy disclosure and copyright infringement. Chatbot companies and users may pursue their interests in the moment but overlook the long-term adverse effects that the pursuit of these interests can have on the information market. All three of these risks represent a type of market failure that may warrant some forms of regulatory intervention.

### 1.   *Re Internal-Market Competition*

In this Article, the discussion up until now has shown that the current chatbot market is competitive and that users, if not satisfied with their current chatbot, can switch to alternative chatbots or multi-home in different systems.[195] However, this impressive degree of competitiveness should not lead policymakers to view the generative AI industry in a carefree manner. Primary concerns are twofold: the increasing market concentration that we can predict for the future market and the potential anticompetitive practices engaged by incumbents seeking to entrench their already impressive market positions. Both patterns may indirectly dampen the factors that incentivize chatbot companies' improvement of their output quality.

---

[193] *Cf.* Liebeck v. McDonald's Restaurants, P.T.S., Inc., No. CV-93-02419, 1995 WL 360309 (N.M. Dist. Ct. Aug. 18, 1994).

[194] *See* discussion *infra* Section IV.B.2.

[195] *See* discussion *supra* Section III.B.1.

It has been found that economies of scale and economies of scope in the generative AI industry may, for the foreseeable future, increase the market concentration of incumbents.[196] This risk stems primarily from the extremely high fixed costs of deploying and developing foundation models and the much lower variable costs of, for example, fine-tuning the models and applying them to various fields.[197] In addition to inherent market structures that favor incumbents and block the entrance or growth of would-be competitors, incumbents may protect their market position by adopting such supplementary strategies as vertical integration, self-preferencing and discrimination, predatory pricing, collusion, and the creation of lock-in effects.[198]

Given the potential anticompetitive harm that the above structures and practices can have on current and prospective market players, some *ex-ante* regulatory interventions may be indispensable. Two articles have made arguably outstanding contributions in this direction, namely *Market Concentration Implications of Foundation Models* by Vipra and Korinek,[199] and *An Antimonopoly Approach to Governing Artificial Intelligence* by Narechania and Sitaraman.[200] As Narechania argues, "[w]hile AI might be new, the problems that arise from concentration in core technologies are not."[201] Correspondingly, they have proposed several antimonopoly tools that might alleviate concerns about market concentration and that include several old-fashioned regulatory strategies directly copied from regulations previously applied to telecommunications and digital platforms, such as structural separations, nondiscrimination, open access, rate regulation, and interoperability.[202]

---

[196] *See* Jai Vipra & Anton Korinek, *Market Concentration Implications of Foundation Models: The Invisible Hand of ChatGPT*, BROOKINGS (Center on Regulation and Markets Working Paper No.9, Sept. 2023), https://www.brookings.edu/articles/market-concentration-implications-of-foundation-models-the-invisible-hand-of-chatgpt/ [https://perma.cc/7PDF-67PB].

[197] *See id.* at 9–12.

[198] *See id.* at 23–29; *see also* Tejas N. Narechania & Ganesh Sitaraman, *An Antimonopoly Approach to Governing Artificial Intelligence*, (Jan. 19, 2024) (manuscript at 23–29) (unpublished manuscript) (on file with SSRN), https://papers.ssrn.com/abstract=4597080 [https://perma.cc/K4JC-B6D3].

[199] *See* Vipra & Korinek, *supra* note 196.

[200] *See* Narechania & Sitaraman, *supra* note 198.

[201] Tejas Narechania, *A Path to Regulating AI*, BERKELEY L. (Jan. 19, 2024), https://www.law.berkeley.edu/sidebar/tejas-narechania-regulating-ai-technology/ [https://perma.cc/PBJ2-YBFT].

[202] *See* Narechania & Sitaraman, *supra* note 198, at 44–47; Vipra & Korinek, *supra* note 196, at 30–36.

It is noteworthy that these regulatory proposals are not content-based, but market-based and technology-neutral. They target market structures and should be evaluated dynamically within the context of the entire industry, which can be considered from the perspective of the two-decade policy shift affecting network neutrality rules in the United States.[203] Although some regulatory proposals may face stiff opposition and may indeed prove unnecessary, the effort to design *ex-ante* regulations that invigorate market competition is, at the very least, a push in the right direction.[204]

### 2. Re External-Market Competition

Privacy disclosure and copyright infringement are two risks that neither internal- nor external-market competition can adequately address, because some information consumers may not be rational enough and may welcome privacy disclosure and copyright-infringing output.[205] To fulfill these needs, chatbot companies may fail to take sufficient steps to counter these risks.[206] Given such market failure, governments may need to consider some properly tailored regulations.

#### a. Privacy Risks: Personalized Protection Tailored to Users' Individual Expectations

One proposal to mitigate privacy risks is to require that chatbot companies access only publicly available data.[207] But this proposal only addresses one part of the problem. While it truly would go far in immediately and effectively mitigating privacy risks, some aspects of this proposal require clarification and moderation. First, the definition of "public data" merits detailed elaboration. In fact, the standards that determine what qualifies as public data can be ambiguous and quite subjective. The flip side of this issue—"what constitutes privacy?"—is also highly subjective.[208] A unitary standard for privacy protection creates either false positives when individuals do not view their data as private or false negatives when individuals are more privacy-sensitive than the regulators anticipate. Thus, there is little sense in assertions that Wikipedia and other encyclopedias, newspapers, and magazines are publicly intended data while social-media content and public posts created by individuals fall under the category of non-public data.[209] Obviously, some posts on X (formerly Twitter), for example, are in the public realm because users expect their posts to circulate as far as possible. Thus, chatbot companies may find that an initially satisfactory set of standards for making "public–private" distinctions is highly ambiguous. Second, data-curation requirements, such as those which I discussed above, are onerous for AI companies and even detrimental to the whole generative AI industry.[210] AI companies cannot tag each piece of data based on a public–private dichotomy.

---

[203] *See supra* note 44 and accompanying text.

[204] *See* U.S. AI Order, *supra* note 147, at 75208–09 (urging agencies to "promote competition in AI and related technologies, as well as in other markets [by] addressing risks arising from concentrated control of key inputs, taking steps to stop unlawful collusion and prevent dominant firms from disadvantaging competitors, and working to provide new opportunities for small businesses and entrepreneurs"); *see also Readout of White House Meeting on Competition Policy and Artificial Intelligence*, THE WHITE HOUSE (Jan. 20, 2024), https://www.whitehouse.gov/briefing-room/statements-releases/2024/01/20/readout-of-white-house-meeting-on-competition-policy-and-artificial-intelligence/ [https://perma.cc/BJ5J-XMVC].

[205] *See* discussion *supra* Section III.B.2.

[206] *Id.*

[207] *See* Winograd, *supra* note 117, at 649. Here, it is noteworthy that the author argues that the proposal to restrict the purview of training data to "publicly-intended" data is non-mandatory. If so, however, the proposal would become meaningless because chatbot companies have no incentives to limit their training data while regulators cannot enforce such rules.

In light of these concerns, an alternative proposal is a personalized and decentralized privacy-protection scheme: personalized because a one-size-fits-all approach cannot satisfy the heterogeneous privacy expectations and other needs of all people;[211] and decentralized because users, not chatbot companies, should have the final say on whether a piece of data can be disclosed and used. To that end, users who have an opportunity to share their information in a digital setting should be entitled and required to establish their privacy expectations. For example, when X users share a post, X should require that they first confirm if it can serve as training data for LLMs.

Here, a hierarchical approach to privacy protection based on data, recipients, and purposes is very promising and well-suited for LLM data training.[212] Users of a digital service can evaluate whether its privacy policy aligns with their privacy expectations. To perform this evaluation, users need to know what, if any, personal information (data) is shared with specific parties (recipients) and for what specific objective (purpose). Once in possession of this knowledge, users should be able to personalize their privacy settings. Thereafter, all relevant parties, including above all the digital service provider, should adhere to these settings. Picking up on the previous example of X, we should expect that this social media giant would process its collected data according to users' settings and would employ technological and legal resources to prevent LLMs from accessing data that users have identified as being off limits. In this way, LLMs can access a sizable amount of training data without bearing onerous compliance costs related to privacy protection, and simultaneously, users will have an incentive to share information that can serve the training needs of LLMs.

---

[208] *See* Pamela J. Wisniewski et al., *Making Privacy Personal: Profiling Social Network Users to Inform Privacy Education and Nudging*, 98 INT'L J. HUMAN-COMPUT. STUD. 95, 106 (2017) (empirically showing "the complex, multi-dimensional nature of end users' varying privacy behaviors and levels of privacy feature awareness"); Alfred Kobsa, *Tailoring Privacy to Users' Needs*, *in* USER MODELING 2001 303, 303 (Mathias Bauer et al. eds., 2001) (arguing that "a uniform solution for privacy demands does not exist since both user preferences and legal stipulations are too heterogeneous").

[209] *See* Winograd, *supra* note 117, at 650. For relevant debates on whether the information users share on social media is public or not, see, for example, Facebook, Inc. v. BrandTotal Ltd., 499 F. Supp. 3d 720, 739 (N.D. Cal. 2020); HiQ Labs, Inc. v. LinkedIn Corp., 31 F.4th 1180, 1189–90 (9th Cir. 2022).

[210] *See* discussion *supra* Section IV.A.3.

[211] *See, e.g.*, Christoph Busch, *Implementing Personalized Law: Personalized Disclosures in Consumer Law and Data Privacy Law*, 86 U. CHI. L. REV. 309 (2019); *see also* Daniel J. Solove, *Introduction: Privacy Self-Management and the Consent Dilemma*, 126 HARV. L. REV. 1880 (2013) (introducing privacy self-management, consent dilemma, and corresponding solutions).

[212] *See* Yuan Hong et al., *A Hierarchical Approach to the Specification of Privacy Preferences*, INNOVATIONS IN INFO. TECH. (IIT) 660, 660 (2007).

### *b.   Copyright Risks: Softening the Competitive Relationship between Chatbot Output and Copyrighted Material*

Regarding the copyright risks associated with chatbots, my core suggestion is to soften the competitive relationship between chatbot-generated output and copyrighted material. This suggestion—which stands in contrast to a mandatory clearing of copyrighted material from AI databases—is based primarily on the spirit of the first and fourth factors of fair use. Specifically, the first factor of fair use examines "the purpose and character of the [secondary] use" of a copyrighted work, and the fourth factor focuses on "the effect of the use upon the potential market for or value of the copyrighted work."[213] One aspect that these two factors share is their stress on the competitive relationship between secondary creation and copyrighted material.[214] The concept and practice of fair use would suffer gravely if the law permitted the secondary use of a copyrighted work to supersede the work and deprive it of its otherwise deserved market share.[215] Therefore, it is essential to know whether the secondary work is in direct competition with the copyrighted work in an information market, and as such, might silence or supersede the copyrighted one.

---

[213]  17 U.S.C. § 107.

[214]  For the application of factor one, see Campbell v. Acuff-Rose Music, Inc., 510 U.S. 569, 579 (1994) (emphasis added) (citations omitted) ("The central purpose of this investigation is to see . . . whether the new work merely '*supersede[s]* the objects' of the original creation . . . or instead adds something new, with a further purpose or different character, altering the first with new expression, meaning, or message."). For the interpretation of factor four, see Harper & Row, Publishers, Inc. v. Nation Enterprises, 471 U.S. 539, 568–69 (1985) (emphasis added) (citation omitted) ("[T]o negate fair use one need only show that if the challenged use 'should become widespread, it would adversely affect the *potential market* for the copyrighted work.' [ . . . ] [A] fair use doctrine that permits extensive prepublication quotations from an unreleased manuscript without the copyright owner's consent poses substantial potential for damage to the marketability of first serialization rights in general.").

[215]  *See, e.g.*, Andy Warhol Found. for the Visual Arts, Inc., v. Goldsmith, 598 U.S. 508, 528 (2023) (emphasis added) (citation omitted) ("[T]he first factor relates to the problem of substitution—copyright's bête noire. The use of an original work to achieve a purpose that is the same as, or highly similar to, that of the original work is more likely to *substitute for, or 'supplan[t],' the work*.").

In this sense, AI output might be found eligible for fair use protection if chatbot companies adopt sufficient measures to soften the competition between chatbot-generated information and copyrighted material. Here, *Authors Guild v. Google* serves as a very fitting and strong precedent for reference.[216] In this case, the Second Circuit Court of Appeals considered the features of Google's snippet view and, citing the first factor of fair use, ruled in favor of Google.[217] The court explained that the snippet-view function is transformative in that it presents "the searcher just enough context surrounding the searched term to help her evaluate whether the book falls within the scope of her interest (without revealing so much as to threaten the author's copyright interests)."[218] In examining the fourth factor, the court held that, although the snippet function may create a substitute for a copyrighted work, any reduction in the work's market value is so minimal that the fourth factor does not favor the plaintiffs.[219] In other words, "the possibility, or even the probability or certainty, of some loss of sales does not suffice to make the copy an effectively competing substitute." [220] Therefore, the court concluded the following:

> Even if the snippet reveals some authorial expression, because of the brevity of a single snippet and the cumbersome, disjointed, and incomplete nature of the aggregation of snippets made available through snippet view, we think it would be a rare case in which the searcher's interest in the protected aspect of the author's work would be satisfied by what is available from snippet view, and rarer still—because of the cumbersome, disjointed, and incomplete nature of the aggregation of snippets made available through snippet view—that snippet view could provide a significant substitute for the purchase of the author's book.[221]

The lessons drawn from *Authors Guild v. Google* can be generalized into two suggestions useful for both government policymaking and private-sector internal compliance. First, chatbot-generated content should be as transformative as possible.[222] To comply with this rule, chatbot companies can train their LLMs to provide users only summaries or brief excerpts of copyrighted works, never the entire work or lengthy verbatim segments of the work.[223] Second, chatbot-generated content should cite the sources of the output. This suggestion, which satisfies the traceability requirement,[224] enables users to find and use the original source, in turn promoting the circulation of copyrighted material.[225] The combination of transformative chatbot output and heavily cited chatbot output not only greatly reduces any risk to copyright holders, but also grants them a degree of positive publicity from which they could benefit reputationally and financially.[226]

---

[216] Authors Guild v. Google, Inc., 804 F.3d 202 (2d Cir. 2015).

[217] *Id.* at 217–18.

[218] *Id.* at 218.

[219] *Id.* at 224.

[220] *Id.*

[221] *Id.* at 224–25 (emphasis omitted).

These two suggestions should, if implemented, soften the competitive relationship between chatbot output and copyrighted material, and should do so in a way that favors the finding of fair use. For example, suppose ChatGPT responds to users' questions by referring to a copyrighted New York Times article. In this scenario, ChatGPT would more likely be compliant with fair-use standards if it provided only a summary of or brief excerpt from the article, including a precisely cited reference and even a hyperlink to the original source. These approaches can ensure the free flow of diverse chatbot-generated information to users in ways that comply with fair-use standards, promote copyrighted works, and avoid substitution and free-rider problems.

## V. CONCLUSION

Can inter-informational competition liberate generative AI from government regulations targeting chatbot output? This is a question that should be considered and answered by policymakers before they issue ambitious regulatory proposals promising to stamp out the risks posed by generative AI. However, the recent push to establish such regulation has relegated this question to the sidelines. In their reflexive embrace of regulation, researchers and policymakers mistakenly regard all theorized AI risks as unequivocally real, present, and harmful, and seek to control these risks by counterproductively employing direct, heavy-handed regulatory tools that create more adversity than they overcome.

---

[222] *See* Andy Warhol Found. for the Visual Arts, Inc., v. Goldsmith, 598 U.S. 508, 529 (2023) ("The larger the difference, the more likely the first factor weighs in favor of fair use. The smaller the difference, the less likely.").

[223] *Cf. Authors Guild*, 804 F.3d at 224 (emphasis added) ("Snippet view, at best and after a large commitment of manpower, produces discontinuous, tiny fragments, amounting in the aggregate to *no more than 16%* of a book. This does not threaten the rights holders with any significant harm to the value of their copyrights or diminish their harvest of copyright revenue.").

[224] *See* discussion *supra* Section IV.A.5.

[225] *See generally* David Fagundes, *Market Harm, Market Help, and Fair Use*, 17 STAN. TECH. L. REV. 359, 378–85 (2014) (presenting four cases in which unauthorized use of copyrighted works prove beneficial to copyright owners, namely recognition, affirmation, reincarnation, and innovation).

[226] *Id.* at 378–80.

The ideas that I have set forth in this Article challenge the prevailing regulatory fervor over generative AI and redirect these misplaced energies toward a far more rational set of steps. Specifically, I recommend a market-centered framework for evaluating not only the widely recognized risks of chatbot output, but also the most commonly proposed regulatory responses.

This framework rests on four lessons drawn from the long regulatory history of radio, television, broadband internet, and other information technologies. First, policymakers and researchers have a great tendency to overstate the risks and uncertainties caused by emerging technologies and to reflexively embrace harsh conduct-based regulations for the mitigation of potential harms. Second, policymakers and researchers have historically regarded competition as primary and regulation as ancillary. Third, policymakers should narrowly tailor generative AI regulations so that they invigorate the market rather than weaken it. And finally, policymaking and judicial rulings for information technologies should always serve to enhance information consumers' right to access diverse and competing information outlets.

Resting on these four foundational lessons, the proposed framework should require policymakers to evaluate the *internal market competition* between a chatbot and another chatbot (*e.g.*, ChatGPT vs. Claude) and the *external market competition* between a chatbot and other types of information outlet (*e.g.*, ChatGPT vs. Google's search engine) to determine whether the market mechanism can mitigate the risks of chatbot output. I have argued that current inter-informational competition is sufficient to mitigate some risks of chatbot output because chatbot companies themselves, under the market forces, have an inherent incentive to improve their output quality and derisk their systems, even in the absence of any governmental intervention. Information consumers are critical and can switch to an alternative source if they find the current chatbot output unreliable. Some empirical findings also demonstrate that most chatbot-content risks are overblown and can be mitigated by largely unfettered inter-informational competition. However, self-regulation of the information marketplace may fail to mitigate privacy disclosure and copyright risks. Users and chatbot companies are sometimes shortsighted and care only about satisfying their current interests. Thus, to ensure the longer-term health of the information ecosystem, policymakers should consider some proportionately tailored regulations.

Given how powerful the information market is at mitigating chatbot content risks, we should ask ourselves: Whether a specific regulatory proposal is necessary and proportionately tailored? My position is that some proposals, such as mandatory prohibitions, licensure, curation of datasets, and notice-and-respond mechanisms, are not well-tailored—they are therefore unnecessary in the context of chatbot content regulation. Instead, transparency, traceability, and audit proposals have great potential to enhance inter-informational competition and further alleviate chatbot content risks. Moreover, in legal cases determining a chatbot company's liability, judges should first ascertain the existence of real harm as opposed to mere risk, and second, should calculate an appropriate allocation of duties by performing a risk-utility test (the Hand formula or, more precisely, the marginal Hand formula).

Not all government intervention is unwelcome. This Article has proposed several regulatory suggestions for unresolved issues derived from potential market failures. Concerning concentration tendencies and anticompetitive practices in the internal chatbot market, I have endorsed several *ex-ante* regulations to adjust the competitive relationship among AI companies. So long as they are subject to dynamic market analysis, these market-based rules will nurture healthy competition, which in turn will incentivize improvements in both output quality and system derisking. Regarding privacy disclosure risks, I have suggested a personalized and decentralized approach to protecting user privacy. In this way, policymaking can reach a delicate balance between the need to access sufficient data for LLM training and the protection of users' privacy expectations. As for copyright infringement risks, government regulations should require that chatbot companies train their LLMs to paraphrase, summarize, and cite copyrighted works. These steps will soften the competing and conflicting relationship between chatbot companies and copyright holders and will thereby counter copyright silencing and the free-rider problem.

My central aim in this Article has been to generate a preliminary analytical framework that policymakers can use for the even-handed assessment of regulatory proposals. My expectation is that the framework will help cool down the current regulatory frenzy over the much-hyped content risks attributed to generative AI. The present study can act as a blueprint for the next stage of research in this field. First, the proposed framework can broaden and deepen our approaches to analyzing AI risks, thereby refining the criteria we use when trying to determine what constitutes a good regulation. Second, the proposed framework, given its preliminary nature, invites more research on how to design and employ indirect, market-based regulation to effectively mitigate AI risks. Third, the proposed framework will benefit from empirical research on information market competition: How competitive is the current chatbot market? Does chatbot output face substantial competition from other types of information outlets? And are information consumers critical enough to differentiate the quality of various information outlets, switch to a more reliable outlet if there is one, and hence avoid falling prey to the lock-in effect? Fourth, my suggestions in this Article have a great bearing on the education of information users: how can governments, businesses, and individuals themselves ensure that society is equipped with enough technological expertise and intellectual wherewithal to navigate an information marketplace increasingly flooded with problematic information?